\documentclass[a4paper]{jpconf}
\usepackage[all,poly]{xy}
\usepackage{amssymb,amsmath}
\usepackage{graphicx}
\begin{document}
\title{Projective Ponzano--Regge spin networks and their symmetries}

\author{Vincenzo Aquilanti$^1$, Annalisa Marzuoli$^{2,3}$}

\address{$^1$ Dipartimento di Chimica, Biologia e Biotecnologie, via Elce di Sotto 8, Universit\`a  di Perugia,  06123 Perugia, I}

\address{$^2$ Dipartimento di Matematica `F. Casorati', Universit\`a  degli Studi
di Pavia, via Ferrata 1, 27100 Pavia, I}

\address{$^3$ INFN, Sezione di Pavia, via Bassi 6, 27100 Pavia, I}

\ead{annalisa.marzuoli@unipv.it}

\begin{abstract}
We present a novel hierarchical construction of projective spin networks of the Ponzano-Regge type
from an assembling of five quadrangles up to the combinatorial 4-simplex  compatible with a geometrical 
realization in Euclidean 4-space. The key ingredients are the projective Desargues configuration
and the incidence structure given by its space-dual, on the one hand, and the Biedenharn--Elliott identity for the $6j$ symbol of SU(2), on the other. The interplay between projective-combinatorial and algebraic features relies on the recoupling theory of angular momenta, an  approach to discrete quantum gravity models carried out successfully over the last few decades. The role of Regge symmetry --an intriguing discrete symmetry of the $6j$  which goes
beyond the standard tetrahedral symmetry of this symbol-- will be also discussed in brief to highlight its role in providing a natural regularization of projective spin networks that somehow mimics  the standard regularization through a q-deformation of $SU(2)$.
\end{abstract}

\section{Introduction}
The Hilbert system of  axioms for an incidence geometry deals with ideal elements such as points, lines and planes, together with the notion of concurrence (incidence) relations \cite{Cox,HilCoh}. The resulting pre--metric admissible configurations  can be also studied as concrete (finite) sets of points, lines and planes in specific geometries, the Euclidean, affine and projective planes or their higher--dimensional counterparts: one then speaks of 
{\em realizations} of some given incidence configuration.
The simplest abstract incidence structures
have been exploited in the (re)coupling  theory of $SU(2)$ angular momentum since the earliest times of Fano and Racah \cite{FaRa1959} and, more in general, combinatorial methods
have been widely applied to the study of discretized quantum spacetimes since the seminal contributions of Giorgio Ponzano and Tullio Regge \cite{PoRe} and  Roger Penrose \cite{Pen1971}. `Quantum spin networks', a term  that nowadays includes quite a wide range of interlaced research fields --from basic features and applications of quantum field theories in low dimensions \cite{CaMaRa2009,CaMaLNP}, up to  the `Loop' approaches  to the quantization of General Relativity \cite{RovBook,AsPu2017}-- will be shown to emerge solely from the basic axioms and theorems of finite--dimensional projective geometry suitably combined with quantum angular momentum theory as rooted in the diagrammatic formalism of Yutsis graphs \cite{YuLeVa}
(a self--contained account on projective configurations and 
diagrammatic--combinatorial tools is collected in two appendices).

The combinatorial--projective structures emerging from the hierarchical construction carried out in section \ref{SecSeconda} are  interpreted as associated with
collective modes of  elementary `quantum  geometries' on the gravitational side
\cite{CaCaMa2002,AqMaMaJPA2013,Tesi,AqMaMaJPCS2014}, and as related to integrable quantum many-body systems of interest also in atomic and molecular physics 
\cite{Aquila1,Aquila2,Aquila3}, on the other.
We recognize in particular that --starting from  the $6j$ symbol of $SU(2)$ encoded into  a quadrangle associated with the projective configuration
of four points joined in pairs by six distinct lines-- a recollection of five nested
quadrangles is projectively nothing but
the Desargue configuration $(10)_3$ (ten points and ten lines, with each of its ten points  incident to three lines and each of its ten lines  incident to three points).
We will show how such a configuration can be consistently labeled with $SU(2)$ 
spin variables as to comply with the algebraic content of the Biedenharn--Elliot
(BE) identity treated in section \ref{SecSeconda}. Moreover:
{\em i)}  Desargues configuration  can be  realized  in a 3D Euclidean space, a well known fact ensured by Desargues' theorem 
which holds true in any dimension $>$ 2, see appendix \ref{AppSeconda}; 
{\em ii)} according to the basic axioms of projective geometry,
each theorem must be valid under projective dualities, to be meant  as either the usual (point $\leftrightarrow$ line) exchanges  in  the plane realization or as (point  $\leftrightarrow$ plane) and (line $\leftrightarrow$ line)
exchanges when viewed in 3D space. On the basis of the above remarks we will remind first
how  the `space--dual' of Desargues configuration has a combinatorial content that matches to the realization of an
Euclidean 4-simplex (bounded by five tetrahedra) and then we will provide the natural $SU(2)$ labeling of its edges. This result is relevant for all Ponzano--Regge-like  quantum gravity models proposed over the years \cite{ReWi2000}, since it would explain
the apparent dichotomy in the assignment of spin labels to triads of angular momenta: in Yutsis-type
diagrams (and in most models used in Loop quantum gravity) 
the correspondence is (triad) $\rightarrow$ (3-valent vertex), 
while in Ponzano--Regge and other approaches \cite{CaCaMa2002} it is assumed that 
(triad) $\rightarrow$ (triangular faces) ({\em cf.} appendix \ref{AppPrima}). 
Actually both correspondences are fully consistent in 3D
Euclidean space  because they are related by Poincar\'e 2D duality for polyhedra, and the tetrahedron is obviouly self--dual in this sense. 
However,
if we keep on assuming, much in the spirit of Regge Calculus \cite{Regge1961}, the prominent role of the assignment (edge length) $\rightarrow$ (spin label), then the 4-simplex Desargue spin network automatically complies
with this requirement and inherits triangular inequalities on triangular faces just as a consequence of
the projective dualization procedure. 
(Recall that in the Euclidean path integral formulation of Regge--discretized gravity the  measure is expressed in terms of edge lengths, the discrete analog
of the functional measure over  metrics, the latter being the dynamical variables 
of General Relativity). In the final part of section \ref{SecSeconda} we will 
briefly comment on 
the issues of dimensional reduction  and increase in this upgraded
projective context.

In section \ref{SecTerza} 
the association between the Desargues configuration and the Biedenharn--Elliott  identity for the $6j$ symbol
will be established and worked out  in details 
(this algebraic identity has a number of interpretations and implications in many branches of mathematical physics, and we will briefly review some of them in the concluding remarks). In this section it will be recognized how the BE identity complies with the two projective Desargues spin networks viewed, on the one hand, as a collection
of five nested quadrangles, and as associated to the 4-simplex on the other. Such an algebraic setting 
can be further improved by calling into play Regge symmetry of the $6j$ symbol 
\cite{Regge1959},
previously discovered for the Wigner $3j$ symbol \cite{Regge1958}. The prominent feature of  Regge symmetry is its functional, not geometric origin, grounded into the Racah sum rule for the $6j$ symbol, {\em cf.} \cite{BiLouck9,Russi} for reviews and original references. 
However, over the years it has been recognized as multi--faceted and still intriguing in many respects, ranging from its interpretation as non trivial automorphism group of a set of not-congruent Euclidean tetrahedra \cite{Roberts1999}, up to its relation with Okamoto symmetry of the Painlev\'e VI differential equation \cite{Boa2007}. We  have been working on the aspects related to quadratic operator algebras, a frame that includes 
general (binary and symmetric) recoupling theory of $SU(2)$ angular momenta, as well as
on the semiclassical analysis of three--term recursion relations, {\em cf.} 
\cite{AqMaMaJPA2013,Tesi} also for 
a complete list of references. The definition of Regge symmetry is briefly reviewed at the beginning of the section,
and it will be shown that, by suitably ordering and keeping fixed the  spin labels assigned to the four edges of a given reference quadrangle,  the quantum amplitude of
each Desargues spin network is automatically regularized in terms of a positive integer as happens when working {\em ab initio} within the framework of 
representation theory of the quantum group $SU(2)_q$ at $q$= root of unity 
\cite{BiLohe,Koek2010}.

In section \ref{SecQuarta}  a glimpse to other interconnections and further developments will be addressed.

%%%%%%%%%%%%%%%%%%%%%%%%%%%%%%%%%%%%%%%%%%%%%%%%%%%
%%%%%%%%%%%%%%%%%%%%%%%%%%%%%%%%%%%%%%%%%%%%%%%%
\section{\label{SecSeconda}Desargues spin networks}
%%%%%%%%%%%%%%%%%%%%%%%%%%%%%%%%%%%%%%%%%%%%%%%%%%%%%%%
%%%%%%%%%%%%%%%%%%%%%%%%%%%%%%%%%%%%%%%%%%%%%%%%%%%%%%%%
In this section we are going to present a novel hierarchical construction of projective spin networks,
from an assembling of five quadrangles up to the combinatorial 4-simplex 
(5 vertices, 10 edges, 10  triangular faces, 5 tetrahedra) compatible with a geometrical 
{\em realization} in Euclidean 4-space. We proceed in parallel and consistently with 
`combinatorial' labeling of configurations, on the one hand, and with labeling provided by angular momentum theory, on the other. The algebraic content of the construction, together with an improved regularization
procedure, are addressed in section \ref{SecTerza}.
We refer to  appendices  \ref{AppPrima} and \ref{AppSeconda} for a primer on angular momentum (re)coupling 
theory and for the discussion 
of the projective features of the configurations and diagrams of this section. Notations and conventions are
those of the classic handbook \cite{Russi}. 

%%%%%%%%%%%%%%%%%%%%%%%%%%%%%%%%%%%%%%%%%%%%%%%%%%%%%%%%%%%%%%%
\subsection{\label{SecSecondaA}Combinatorics of nested quadrangles}
%%%%%%%%%%%%%%%%%%%%%%%%%%%%%%%%%%%%%%%%%%%%%%%%%%%%%%%%%%%%%%%%%%%

\begin{figure}[htbp]
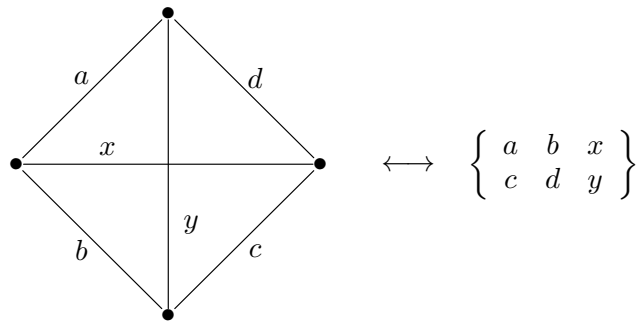

\[
\xy
0*{}="A";
<0cm,2cm>*{\bullet}="B";
<2cm,0cm>*{\bullet}="C";
<4cm,2cm>*{\bullet}="D";
<2cm,4cm>*{\bullet}="E";
"B";"E" **@{-} ?(0.5)*!/_2mm/{a};
"E";"D" **@{-} ?(0.5)*!/_2mm/{d};
"D";"C" **@{-} ?(0.5)*!/_2mm/{c};
"C";"B" **@{-} ?(0.5)*!/_2mm/{b};
"E";"C" **@{-} ?(0.7)*!/_3mm/{y};
"B";"D" **@{-} ?(0.3)*!/_2mm/{x};
<6.5cm,2cm>*{\longleftrightarrow\;\;\;\left\{\begin{array}{ccc}
a & b & x\\
c & d & y
\end{array}
\right\}}
\endxy
\]
\caption{The Yutsis diagram of the $6j$.}
\label{6jYutsis}
\end{figure}

The basic building block of the  construction  is the
$6j$ symbol presented through its Yutsis diagram 
as in Fig.\ref{6jYutsis}
or, equivalently, as the complete quadrangle of Fig.\ref{App4}, where  ($a,b,c,d,x,y)$ $\in \{0,1/2,1,3/2,\dots \}$
label irreducible representations ({\em irreps}) of the Lie group $SU(2)$. 

\begin{figure}[h]
\centering
\begin{minipage}{17pc}
\includegraphics[width=17pc]{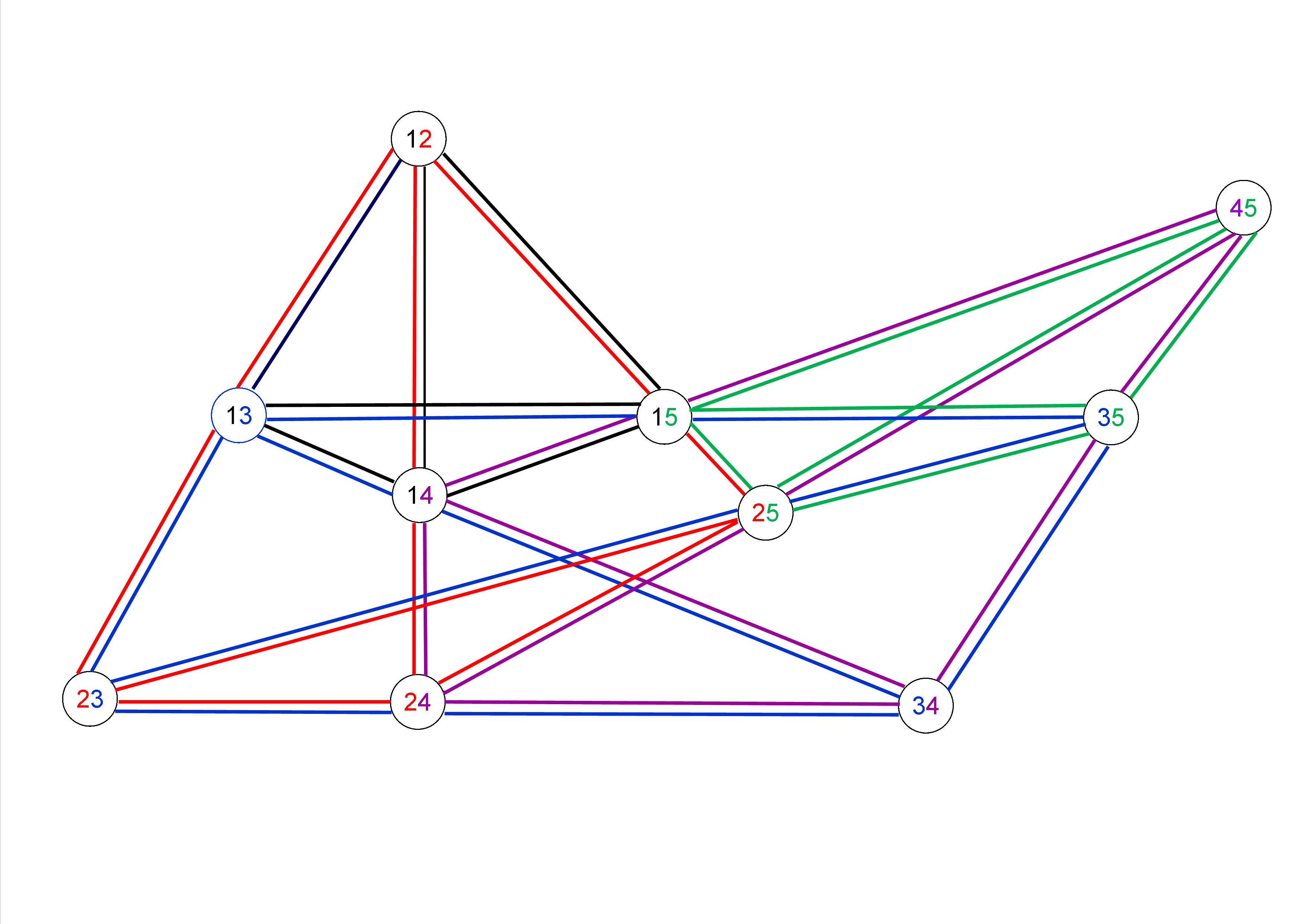} 
\caption{\label{Dbolle}Combinatorial labeling of Desargues configuration $(10)_3$.}
\end{minipage}\hspace{3pc}%
\begin{minipage}{17pc}
\includegraphics[width=17pc]{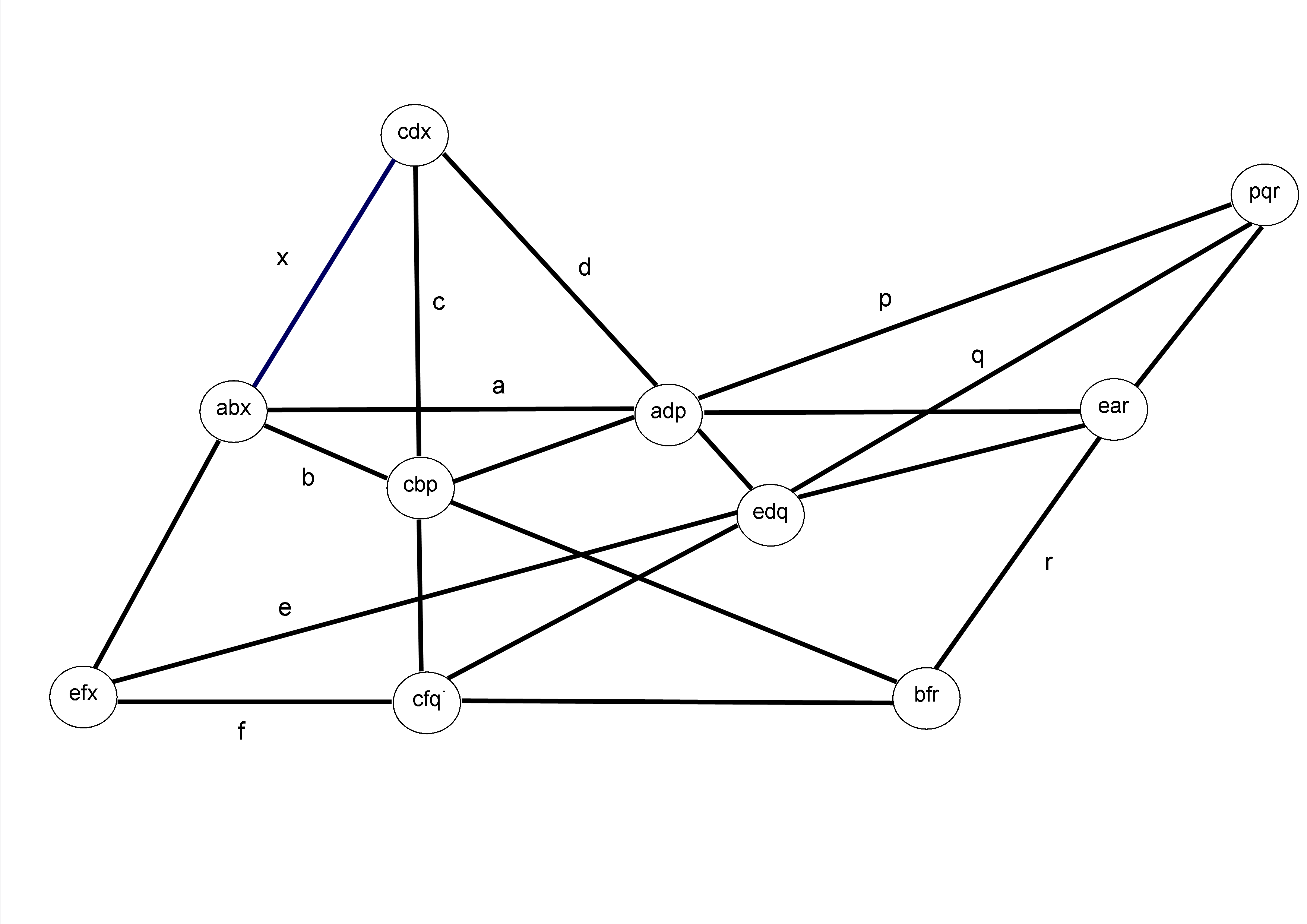}
\caption{\label{Dtriadi}Spin labeling of Desargues configuration $(10)_3$.}
\end{minipage} 
\end{figure}

The Desargues configuration $(10)_3$  shown in Fig.\ref{Dbolle} --the symmetric incidence structure made of ten points and ten lines-- turns out to be combinatorially
consistent with an assembling of five quadrangles, $\mathcal{Q}_1$, $\mathcal{Q}_2$, $\mathcal{Q}_3$, 
$\mathcal{Q}_4$ and $\mathcal{Q}_5$,
in one--to--one correspondence with the coloring (Black, Red, Blue, Purple, Green).
The ten bubbles which stand at the intersection of three lines are 3-valent vertices and 
have a double (unordered) labeling since each of them is shared by two quadrangles. 
The six edges of each quadrangle inherit colors from (one of) the colors of vertices.
The angular momentum (spin) labeling of the five quadrangles --to be associated with the five $6j$ symbols of the BE identity in the following section-- is depicted in 
Fig.\ref{Dtriadi} and is 
given by the correspondences
\begin{equation}\label{cinque6j}
\mathcal{Q}_1
\leftrightarrow 
\begin{Bmatrix}
  a & b & x\\
c & d & p
 \end{Bmatrix}; 
 \mathcal{Q}_2
\leftrightarrow 
\begin{Bmatrix}
  c & d & x\\
e & f & q
 \end{Bmatrix};
 \mathcal{Q}_3
\leftrightarrow 
\begin{Bmatrix}
  e & f & x\\
b & a & r
 \end{Bmatrix}; 
  \mathcal{Q}_4
\leftrightarrow 
\begin{Bmatrix}
  p & q & r\\
f & b & c
 \end{Bmatrix};
  \mathcal{Q}_5
\leftrightarrow 
\begin{Bmatrix}
  p & q & r\\
e & a & d
 \end{Bmatrix}. 
\end{equation}
 
It is worth pointing out that now  the incidence structure underlying Desargues configuration
has to be thought of as {\em geometrically} realized in a $3D$ projective or Euclidean space (namely the perspective point of Desargues' Theorem must be in general position with respect to the planes of the two triangles, 
{\em cf.}  Fig.\ref{App7}). In a $3D$ Euclidean space, in particular,  the five quadrangles considered so far would  appear nested into each other,  and not glued to each other to form  a polyhedron. However,  the duality principle (appendix \ref{AppSeconda}) can be called into play to establish the actual existence of the space--dual of
 Desargues configuration,  which will be proven below to be compatible not only with a combinatorial 4-simplex as noted in \cite{HilCoh,Barnes}, but also with a spin  labeling of the 4-simplex.

%%%%%%%%%%%%%%%%%%%%%%%%%%%%%%%%%%%%%%%%%%%%%%%%%%%%%%%%%%%%%%%%%%%%%%%%%%
\subsection{\label{SecSecondaB}Space--dual of Desargues configuration and the  4-simplex}
 %%%%%%%%%%%%%%%%%%%%%%%%%%%%%%%%%%%%%%%%%%%%%%%%%%%%%%%%%%%%%%%%%%%%%%%%%%%%%%%%%%%%%%

\begin{figure}[h]
\centering
\begin{minipage}{17pc}
\includegraphics[width=17pc]{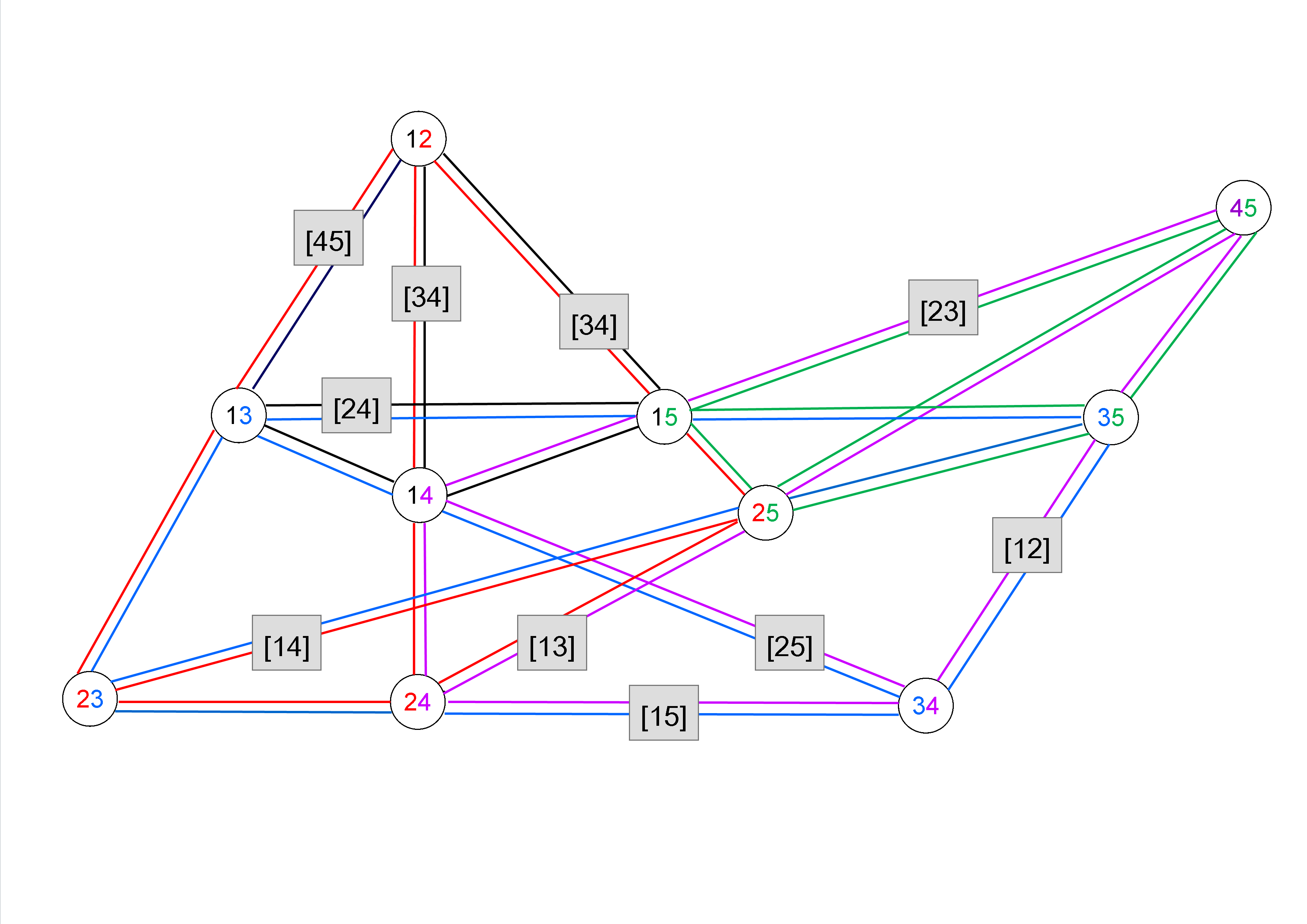}  
\caption{\label{Drette} Dual labeling of lines  put on top of  Desargues configuration of Fig.\ref{Dbolle}.}
\end{minipage}\hspace{3pc}%
\begin{minipage}{17pc}
\includegraphics[width=17pc]{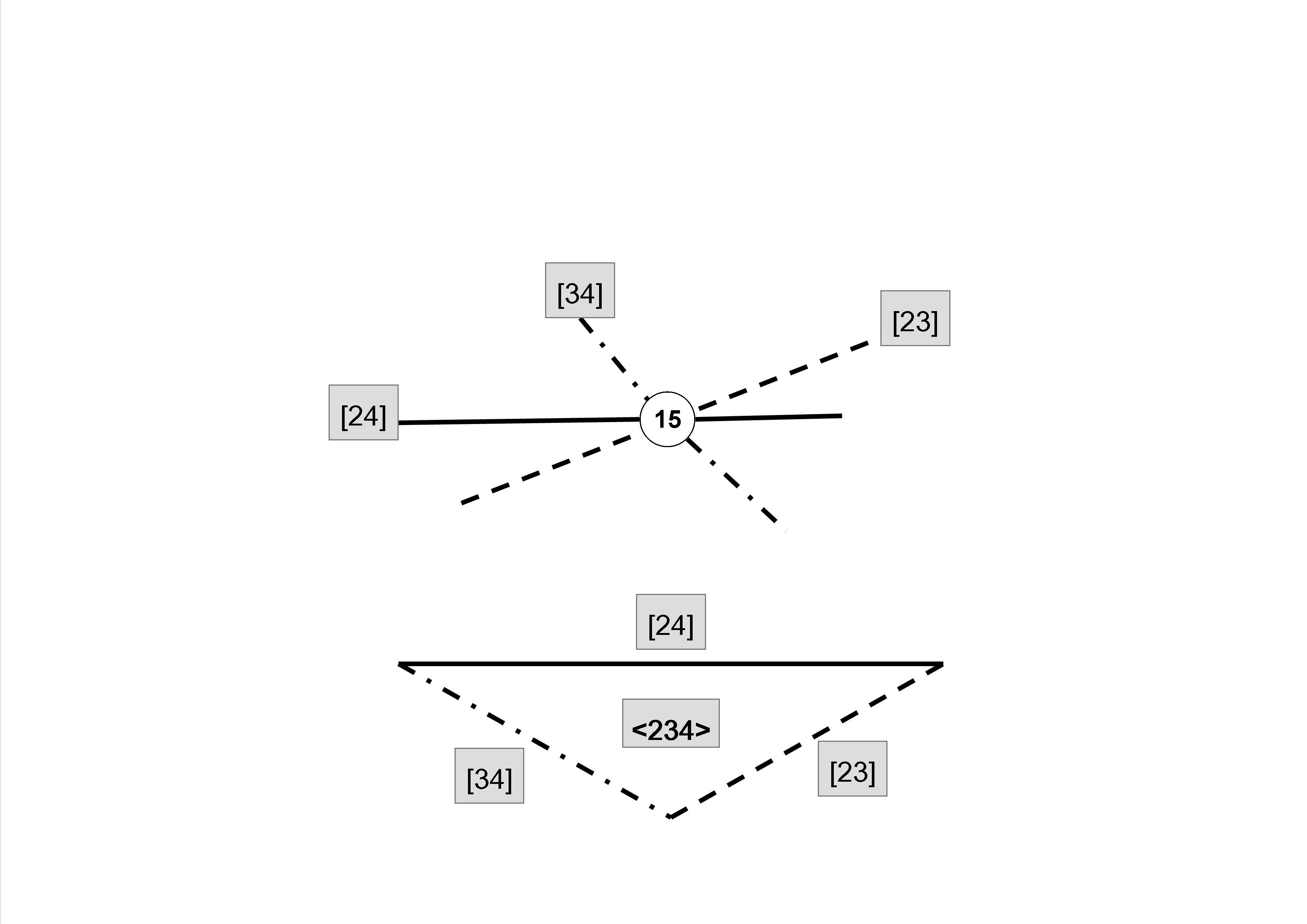} 
\caption{\label{Dscoppio} A blown-up 3-valent vertex and its dual triangle.}
\end{minipage} 
\end{figure}

The notion of duality applies in a straightforward way  to the combinatorial labeling
used in Fig.\ref{Dbolle}. The points are two-colored, denoted by circles or round brackets, $(12)$,
$(23)$, $\dots$; quadrangles are listed  $1,2,3,4,5$ as in Eq.\eqref{cinque6j}; lines are enumerated by
triples of points, {\em e.g.} $(12)$ $(13)$ $(23)$. An  intermediate step consists in 
switching the line labeling to the dual one, where for instance the previous line is
labeled with the two complementary colors $[45]$ within  square brackets, see all the resulting tags
in Fig.\ref{Drette}.

Recall from the final part of appendix \ref{AppSeconda} that in a projective space $P^3$ there is duality between  subspaces of dimension $k$
and $(3-k-1)$ and then the basic correspondences for going through  the space--dual of  Desargues configuration 
(theorem) are 
\begin{quote}
 point $\; \longrightarrow \;\;$ plane,\\
 line $\; \longrightarrow \;\;$ line,\\
 quadrangle $\; \longrightarrow \;\;$ point,
 \end{quote}
where the last pairing  can be worked out starting from the dualization of  labeling of
the original quadrangles, {\em e.g.} $\mathcal{Q}_1$ $\rightarrow$ vertex $\{2345\}$ (curly brackets)
in the space--dual configuration. 
Moreover, noting that  the intersections of pairs of quadrangles represent the vertices of the original configuration of Fig.\ref{Dbolle}, namely
\begin{equation}\label{IntQ}
\mathcal{Q}_i\, \cap \mathcal{Q}_j\,= (ij)\,, \;\;\; i,j=1,2,3,4,5\, \;\,\text{with}\,i \neq j\,,
\end{equation}
and recalling that any such  vertex is actually the intersection of three lines 
(a {\em triad} of spin labeling according to the dictionary of   Fig.\ref{Dtriadi}), we are forced to improve the  duality
(point $\; \rightarrow \;\;$ plane) as 
\begin{equation*}
 \text{3-valent vertex}\; \longrightarrow \;\; \text{triangle}
 \end{equation*}
and to assign the combinatorial labels according to
 \begin{equation}\label{PRtriangle}
 \text{2-color vertex} (ij)\; \rightarrow \; \text{triangle} <klm>
 \end{equation}
 with $k,l,m$ complementary to $(ij)$, {\em cf.}  Fig.\ref{Dscoppio}.\\

\begin{figure}[h]
 \centering
\includegraphics[scale=0.30]{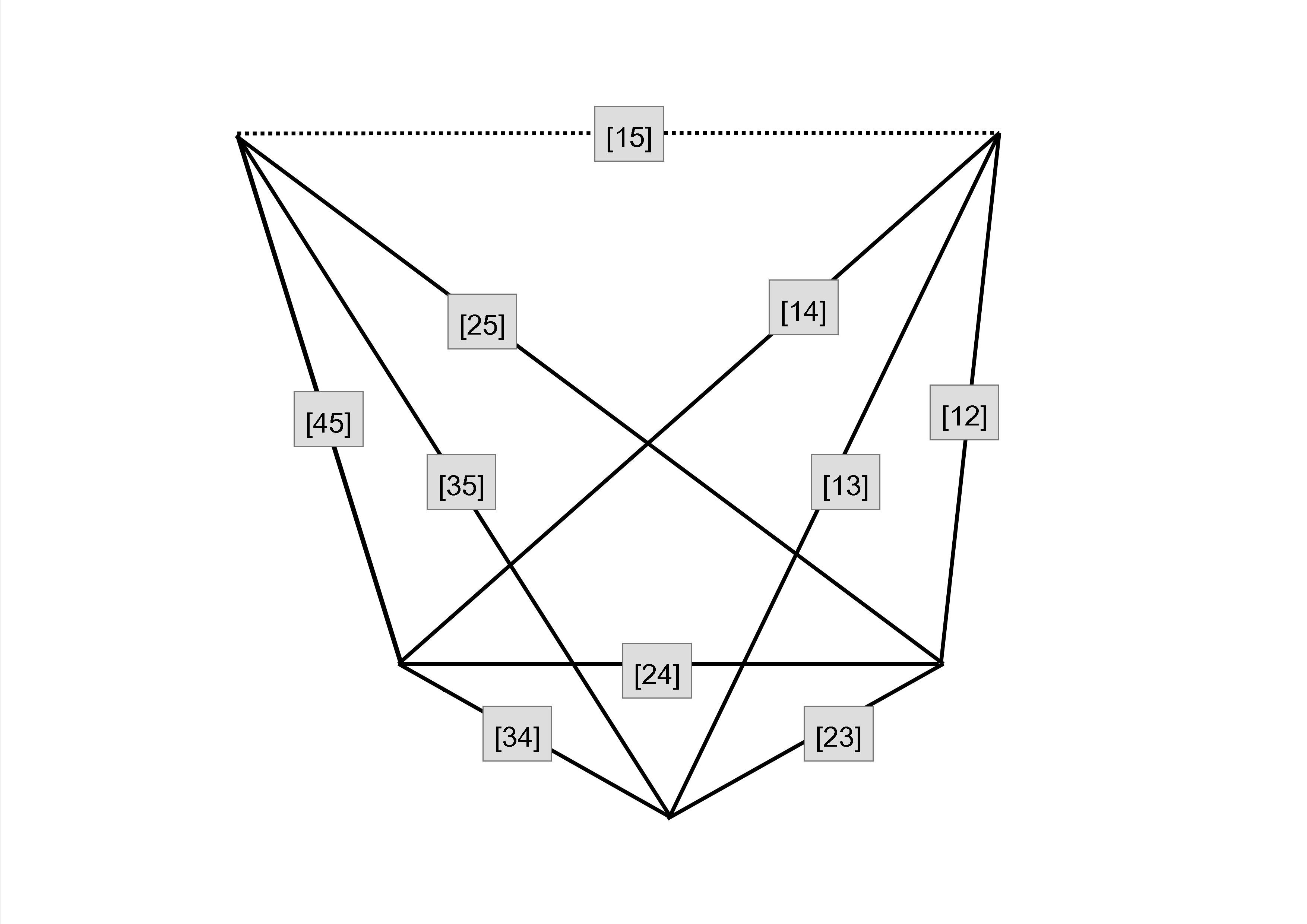} 
\caption{\label{Dtriedi} Space--dual of Desargues' theorem. Here the combinatorial labeling is only on lines
for simplicity, but the labels of the other substructures are easily reconstructed from it:
the two trihedra are incident on the triangle $<234>$ $=$ $ [23] \cup [24] \cup [34]$;
the three planes of each trihedron are spanned by the three (of course not coplanar) lines
 $ [25], [35], [45]$ and $ [12], [13], [15]$, respectively; the theorem ensure the existence of line $[15]$
 and of three planes (triangles) $<125>$, $<135>$, $<145>$. The two vertices on the top  represent the tips of
 the trihedra, labeled $\{2345\}$ and $\{1234\}$ or, by resorting again to complementarity  of labeling,
 denoted $1$ and $5$, respectively.}
\end{figure}

\noindent The statement of the space--dual of Desargues' theorem reads:
\begin{quote}
If two trihedra in 3D space are perspective from a triangle, then there exists 
a line connecting the tips of the two trihedra together with three planes meeting at this line
and passing through the sides of the triangle. 
\end{quote}
The resulting connected ensemble of five vertices, ten lines,
ten triangles and five tetrahedra (actually trihedra, each pair of which intersects a fourth triangle) is combinatorially 
a 4-simplex, denoted $\mathcal{S}_{comb}$ for short. 
The pictorial representation  of this construction is given  in Fig.\ref{Dtriedi} where the quadrangles $\mathcal{Q}_1$ and $\mathcal{Q}_5$ --or, respectively, dual vertices $\{2345\}$ and $\{1234\}$--
have been employed 
consistently with the choice made in Fig. \ref{Dscoppio} (actually any other choice of the
selected pair of quadrangles turns out to be combinatorially  equivalent to each other).
 
 %%%%%%%%%%%%%%%%%%%%%%%%%%%%%%%%%%%%%%%%%%%

As for spin labeling of this new configuration,  observe preliminary  that 
the dual combinatorial labels on edges established in Fig.\ref{Drette} have been  retained
also for lines in the dual Desargues configuration on the basis  of the duality 
(line $\; \rightarrow \;\;$ line) in space. Thus, by resorting to the dictionary of Fig.\ref{Dtriadi},
and denoting  the five tetrahedra of the combinatorial 4-simplex 
$\mathcal{S}_{comb}$ by $\mathcal{T}_i$, the 
correspondence with $6j$ symbols arranged into the upgraded $\mathcal{S}_{spin}$,
is necessarily given by 
\begin{equation}\label{cinque6jbis}
\mathcal{T}_1
\leftrightarrow 
\begin{Bmatrix}
  a & b & x\\
c & d & p
 \end{Bmatrix};
 \mathcal{T}_2
\leftrightarrow 
\begin{Bmatrix}
  c & d & x\\
e & f & q
 \end{Bmatrix} ;
 \mathcal{T}_3
\leftrightarrow 
\begin{Bmatrix}
  e & f & x\\
b & a & r
 \end{Bmatrix}; 
  \mathcal{T}_4
\leftrightarrow
\begin{Bmatrix}
  p & q & r\\
f & b & c
 \end{Bmatrix};
  \mathcal{T}_5
\leftrightarrow 
\begin{Bmatrix}
  p & q & r\\
e & a & d
 \end{Bmatrix}. 
\end{equation}

\begin{figure}[h]
\centering
\includegraphics[scale=0.30]{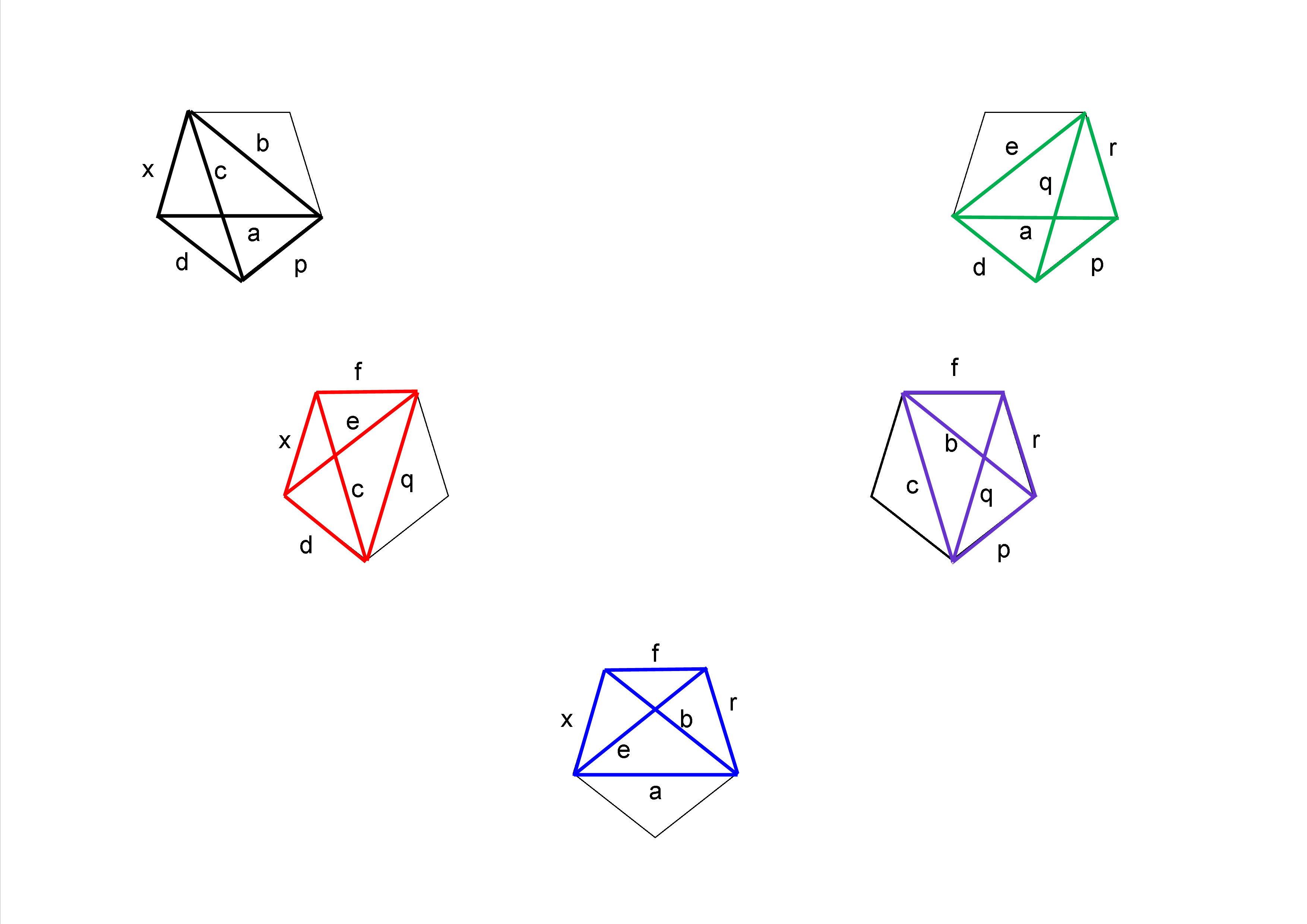} 
\caption{\label{D4sim} Spin labeling of the tetrahedra of the 4-simplex $\mathcal{S}_{spin}$:
($\mathcal{T}_1$, $\mathcal{T}_2$, $\mathcal{T}_3$, $\mathcal{T}_4$, $\mathcal{T}_5$) $\leftrightarrow$ (Black, Red, Blue, Purple, Green), from top left in anti-clockwise order.}
\end{figure}

The crucial difference with respect to  the correspondence between quadrangles of the original Desargues configuration and $6j$'s given in \eqref{cinque6j} consists in recognizing that here the triads of the $6j$ have to be assigned to the triangular faces of tetrahedra as a consequence of  the appropriate {\em projective duality} stated in  Eq. \eqref{PRtriangle}. As  mentioned in appendix \ref{AppPrima} such a presentation of the $6j$ stands at the basis of Ponzano--Regge model for (the semiclassical analysis of) 3D Euclidean gravity, where the spin labels of $SU(2)$ 
{\em irreps} are the weights assigned to edges and give rise to the discrete analog of the 
gravitational path integral measure on the metrics. 
An exploded  view of the space--dual Desargues spin network $\mathcal{S}_{spin}$
is shown in Fig. \ref{D4sim}.  

%%%%%%%%%%%%%%%%%%%%%%%%%%%%%%%%%%%%%%%%%%%%%%%%%%%%%%%%%%%%%%%%%%%%%%%%%
\subsection{\label{SecSecondaC}Geometrical realizations and dimensional reduction and increase}
%%%%%%%%%%%%%%%%%%%%%%%%%%%%%%%%%%%%%%%%%%%%%%%%%%%%%%%%%%%%%%%%%%%%%%%%%%%
As already pointed out at the end of section \ref{SecSecondaA}, the hierarchical construction 
that leads to the Desargues spin network $\mathcal{S}_{spin}$ discussed in section \ref{SecSecondaB}
(complemented with material of appendices \ref{AppPrima} and \ref{AppSeconda})
relies on the assumption that the underlying incidence structures are realized
in a $P^n$ with $n \geq 3$. 
The projective viewpoint provides, on the one hand,  an axiomatic frame 
for dealing with incidence structures 
that {\em emerge} from the simplest configurations of points and lines in the plane,
and an effective procedure based on projective {\em dimensional reductions} to go from 
the geometric realization of a 4-simplex in Euclidean space down to its low--dimensional
cross--sections, on the other.
The lines of reasoning underlying these achievements are summarized in the following steps.
\begin{itemize}
\item 
Any two distinct points are incident with just a line
and  any two lines are incident with at least one point, (Axioms I) and II) of
appendix \ref{AppSeconda}).
The existence of quadrangles (configurations $(4_3\,,6_2)$) is the third  axiom 
of (plane) projective geometry;
\item Among the various finite configurations of points and lines,
Desargues' $(10_3)$ has a special status: in $n=2$ it must be postulated
separately but it is encoded as a theorem for $n \geq 3$, so that
in the latter cases no further axiom is necessary;
\item In $n = 3$ there exist two projective--dual versions of the theorem:
its plane--dual, still denoted   $(10_3)$ because the configuration is symmetric under the 
exchange (points $\leftrightarrow$ lines), and its space--dual, denoted $\mathcal{S}_{comb}$,
discussed at length in section \ref{SecSecondaB};
\item  $\mathcal{S}_{comb}$ can be realized through the projection of an Euclidean 
4-simplex onto  $P^3$;
\item If we slice any such  Euclidean realization of $\mathcal{S}_{comb}$ by an arbitrary solid,
each  line is  cut by this portion of space in a point, each plane in a line and each tetrahedron 
in a plane  so that any cross--section of $\mathcal{S}_{comb}$ 
represents a figure made of ten points, ten lines and five planes, namely 
Desargues' $(10_3)$ \cite{Barnes}.
 \end{itemize}
Obviously the previous remarks touch  basic questions about 
the nature of quantum spacetime, the existence of a fundamental
length  and a number of questions that arise in connection with discretized, 
simplicial or spin networks, approaches to quantum gravity, not to mention 
the issue of dimensional reduction \cite{Car2012}.
Our view is inspired by the (quite elementary) frame provided by projective geometry 
of configurations and by its intriguing relations with angular momentum theory
--relations that have been only partially worked out by other authors,  
sometimes leading to ambiguous conclusions, {\em cf.} the appendices.\\ 
The result we have been proving, 
\begin{quote} the  Desargues configuration and its space--dual are 
respectively cross-sections and projections of an Euclidean 4-simplex,
and
consistent  $SU(2)$ spin labeling can be assigned to each of them, 
\end{quote}
 leaves open the crucial 
question of the existence of a `quantum of space' in (space or spacetime) dimensions D=3,4.
In the light of the previous
considerations, one might either {\em i)} postulate the existence of an Euclidean
4-simplex, assign projective coordinates and go down to a 3D Desargues 
configuration which admits a spin labeling as in Fig.\ref{Dtriadi}, or {\em ii)}
look for an emergent 3D (projective) space from the (re)coupling of angular momenta.
The latter option is actually what occurs according to the combined 
results found by Penrose  \cite{Pen1971} and by Ponzano and Regge \cite{PoRe} which stand at the basis of
all of the spin networks models addressed over the years 
in quantum gravity {\em cf.} \cite{CaMaRa2009,ReWi2000}
for brief accounts and \cite{AsPu2017} for an extended overview
on the Loop approach. Then the rationale of our construction
relies on {\em ii)} and exploits the projective nature of the $SU(2)$ spin networks
up to the emergency of the 4-simplex $\mathcal{S}_{spin}$.

%%%%%%%%%%%%%%%%%%%%%%%%%%%%%%%%%%%%%%%%%%%%%%%%%%%%%%%%%%%%%%%%
%%%%%%%%%%%%%%%%%%%%%%%%%%%%%%%%%%%%%%%%%%%%%%%%%%%%%%%%%%%%%%%%%%
\section{\label{SecTerza}Regge symmetry and improved algebraic setting}
\begin{figure}[h]
\centering
\begin{minipage}{17pc}
\includegraphics[width=17pc]{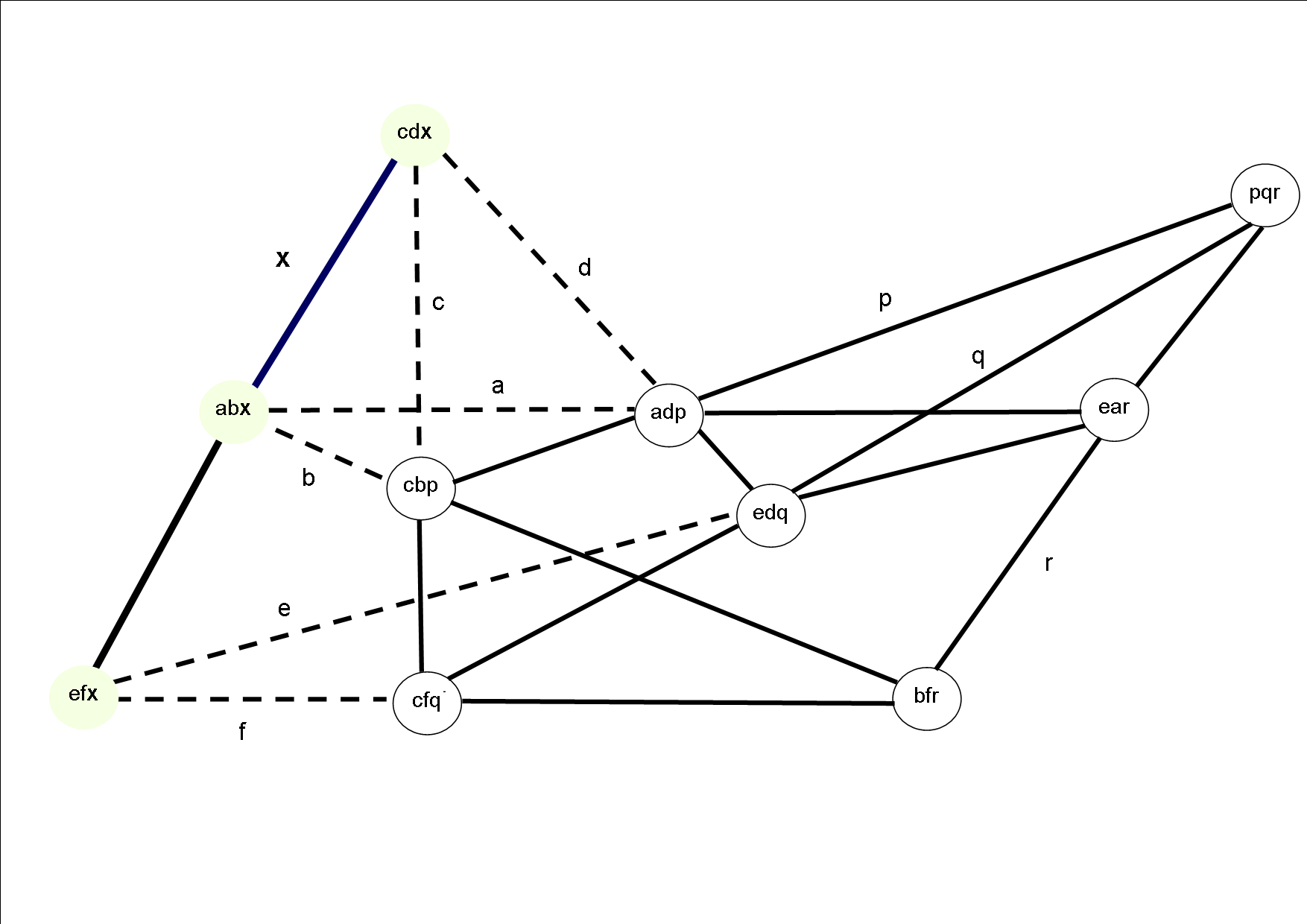} 
\caption{\label{D103BE} Upon summation over $x$ (bold),  three triads 
of spin labels together with the dashed edges disappear:  
only the two quadrangles corresponding to the $6j$'s on the right side of 
the BE identity  survive.}
\end{minipage}\hspace{3pc}%
\begin{minipage}{17pc}
\includegraphics[width=17pc]{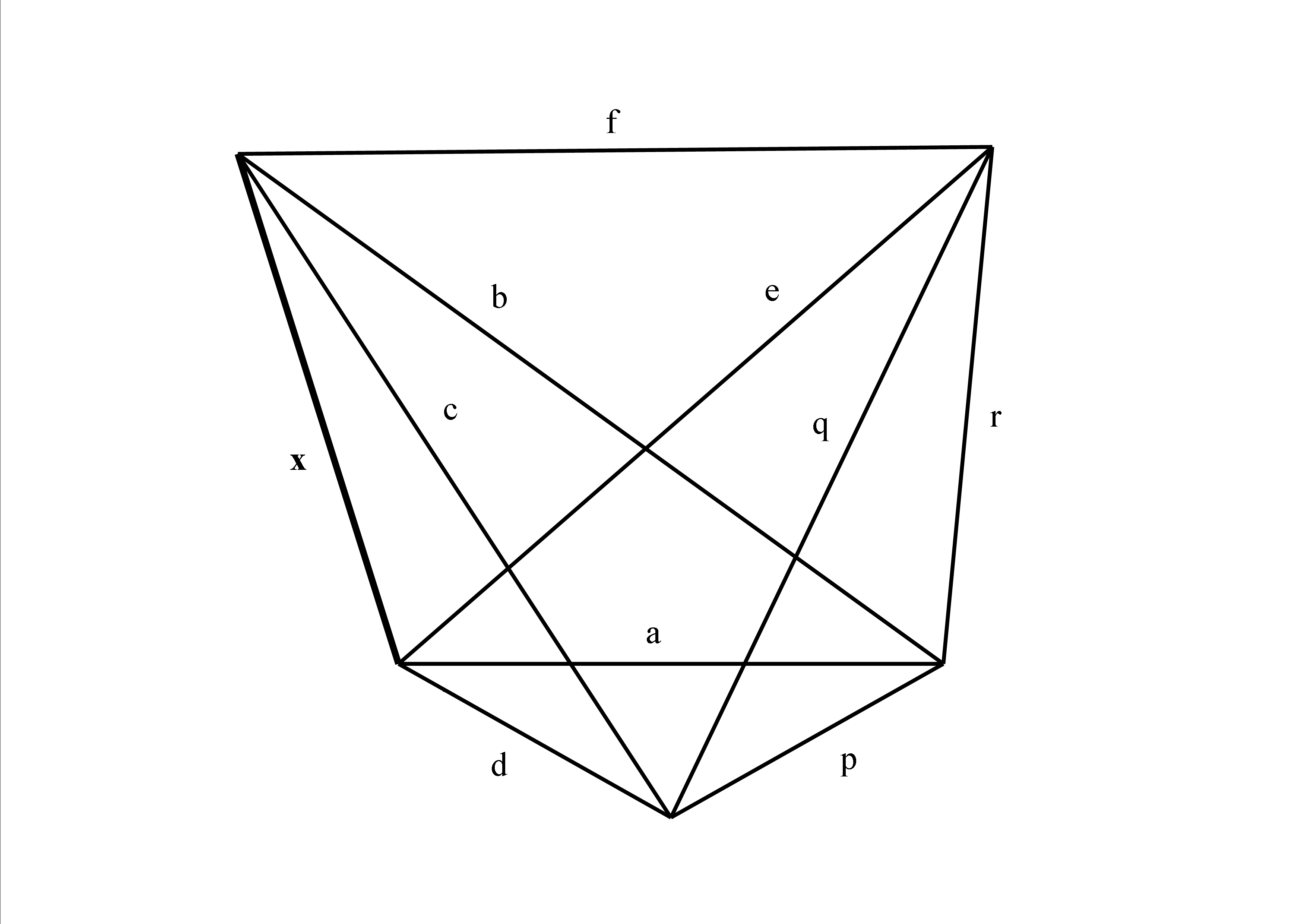} 
\caption{\label{4simBE} The BE identity in the  Desargues space-dual $\mathcal{S}_{spin}$
model: here labelings are identical  to those  of Fig.\ref{D4sim}.}
\end{minipage} 
\end{figure}

 %%%%%%%%%%%%%%%%%%%%%%%%%%%%%%%%%%%%%%%%%%%%%%%%%%%%%%%%%%%%%%%%%%%%%%%%%%%%%
 %%%%%%%%%%%%%%%%%%%%%%%%%%%%%%%%%%%%%%%%%%%%%%%%%%%%%%%%%%%%%%%%%%%%%%%%%%%
 As is well known the defining relations for $6j$ symbols of $SU(2)$ are
 the orthogonality condition,
 \begin{equation}\label{Ort6j}
 \sum_x (2x+1) 
\begin{Bmatrix}
  a & b & x\\
c & d & y
 \end{Bmatrix}
 \begin{Bmatrix}
  c & d & x\\
a & b & y'
 \end{Bmatrix} =
 \frac{\delta_{yy'}}{2y'+1}\delta_{(ady)}\, \delta_{(bcy)},
 \end{equation}
where the symbol $\delta_{(...)}$ constrains the entries to belong to a triad, and the Biedenharn--Elliot (BE) identity 
\begin{equation}\label{BEid6j}
\sum_x (-1)^{\varphi +x} 
\begin{Bmatrix}
  a & b & x\\
c & d & p
 \end{Bmatrix}
 \begin{Bmatrix}
  c & d & x\\
e & f & q
 \end{Bmatrix} 
 \begin{Bmatrix}
  e & f & x\\
b & a & r
 \end{Bmatrix} 
 =\,\begin{Bmatrix}
  p & q & r\\
f & b & c
 \end{Bmatrix}
 \begin{Bmatrix}
  p & q & r\\
e & a & d
 \end{Bmatrix}, 
 \end{equation}
 where $\varphi = a+b+c+d+e+f+p+q+r$. The content of this algebraic
relation, depicted in Fig.\ref{D103BE}, complies with the assignments of spin labels/triads to lines/vertices  --resulting in the association $6j$ $\leftrightarrow$
quadrangles (see Eq.\eqref{cinque6j})-- in the   Desargues configuration 
$(10)_3$ as established in Fig.\ref{Dtriadi}. \\

\begin{figure}[h]
\centering
\begin{minipage}{17pc}
\includegraphics[width=17pc]{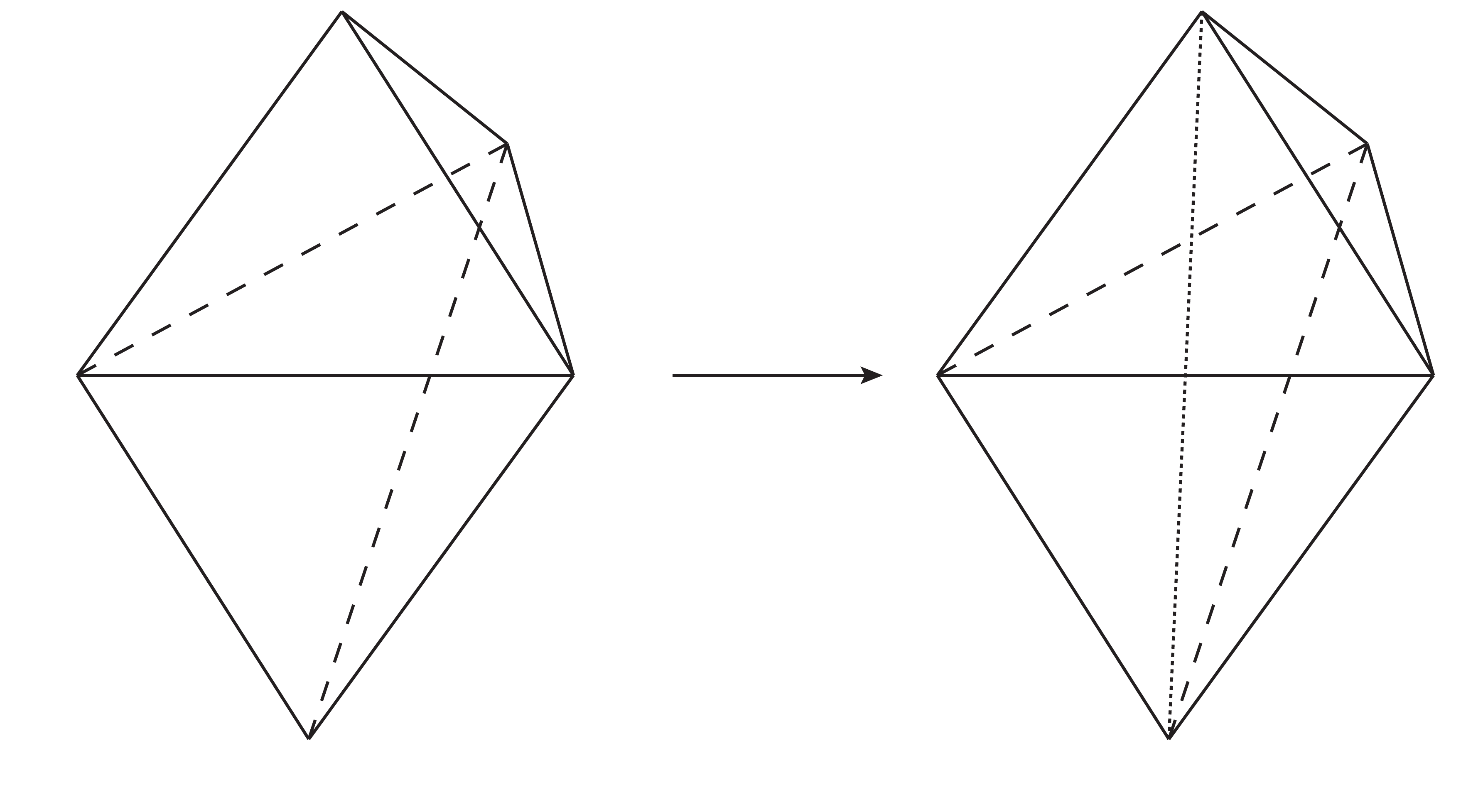}
\caption{\label{App8} Pachner move 2-3}
\end{minipage}\hspace{3pc}%
\begin{minipage}{17pc}
\includegraphics[width=17pc]{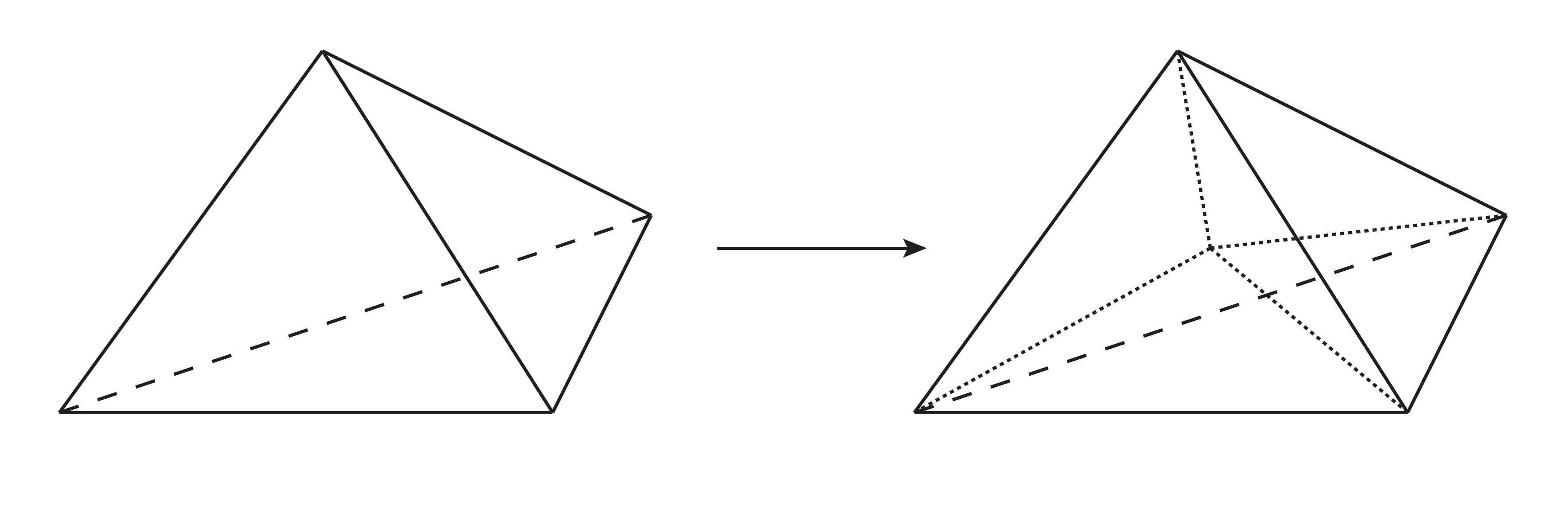}
\caption{\label{App9} Pachner move 1-4}
\end{minipage} 
\end{figure}

\noindent On the other hand, \emph{the same spin assignment} turns out to be compatible
with the 4-simplex frame $\mathcal{S}_{spin}$
associated with the space--dual of Desargues configuration
discussed at length in section \ref{SecSecondaB}, and 
the BE identity holds true consistently,
\emph{cf.}  Fig. \ref{4simBE}. 
In other words, the geometric content of the BE identity in the Ponzano--Regge 
tetrahedral presentation of the $6j$ (see appendix \ref{AppPrima}) is intrinsically projective, as far as
the two arrangements (3 tetrahedra joined along $x$) and  (two tetrahedra joined along
one common face) cannot coexist in an Euclidean 3-space but can be realized
through  $\mathcal{S}_{spin}$ in dimension 4. The latter remark  holds true for the coexistence
of the arrangements (4 tetrahedra joined together and incident on  a common vertex) 
and (1 tetrahedron) which would correspond  to an identity 
easily derived from \eqref{BEid6j} by resorting to \eqref{Ort6j}).
In Fig.\ref{App8} and Fig.\ref{App9} the purely topological versions
 of the so-called Pachner moves \cite{Pach} for simplicial 3-manifolds are depicted.

%%%%%%%%%%%%%%%%%%%%%%%%%%%%%%%%%%%%%%

Looking back at the $6j$ symbol of Fig.\ref{6jYutsis}, the four labels 
$(a,b,c,d)$  --associated with the consecutive sides of a quadrangle, or equivalently of a tetrahedron {\em cf.} \ref{AppPrima}--
are going to be treated as tunable parameters, while $\tbinom{x}{y}$ is thought of as the 
pair of variables corresponding to the matrix indexes of the $6j$ when 
applied to binary coupled states of three $SU(2)$  angular momenta in the space of the total
angular momentum \cite{Russi}.  As is well known the 24 `classical'
symmetries of the $6j$ can be viewed geometrically as associated to the complete tetrahedral group 
(isomorphic to  $S_4$) and leave the (values of the)
symbol invariant under any permutation of its columns or under interchange of
the lower and upper arguments in each of any two columns. 
The Regge symmetries \cite{Regge1959}  are expressed 
as functional relations among parameters and read
\begin{equation}\label{Rsym1}
\begin{Bmatrix}
a & b & x \\
c & d & y 
\end{Bmatrix}\,=\,
\begin{Bmatrix}
s-a & s-b & x \\
s-c & s-d & y 
\end{Bmatrix}\,:=\,
\begin{Bmatrix}
a' & b' & x \\
c' & d' & y 
\end{Bmatrix},
\end{equation}
where $s := (a+b+c+d)/2$ is the semi--perimeter of the quadrangle
and in the last equality the new  set $(a',b',c',d')$ is defined.
The $6j$ associated with the primed entries represents the Regge 
`conjugate' of the original  quadrangle, and of  the
Ponzano--Regge tetrahedron as well.
Combining the symmetries given in (\ref{Rsym1}) with two more sets obtained
by keeping the  other  pairs of opposite entries fixed,  it turns out 
that the total number of symmetries is 144, which equals  the order of
the product permutation group $S_3 \times S_4$. 

In the $6j$s of Eq.\eqref{Rsym1} the range of the running entries is  $x_{max} -x_{min}=$
$y_{max} -y_{min}= 2$ $ min(a,b,c,d,',b',c',d')$ 
on the basis of the choice of a particularly useful ordering for dealing with
the discrete, four--parameter $6j$-function,
 see  \cite{AqMaMaJPA2013,Tesi}.
This construction highlights  in particular `Regge--invariant'
geometric realizations of both quadrangles and  tetrahedra, a symmetry
which naturally extends to the three-term recursion relations which
hold for the $6j$ symbol \cite{Russi,Aquila5}, on the one hand, and for eigenfunctions
of the volume operator, on the other \cite{CaCaMa2002,AqMaMaJPCS2014}.\\
Here we just add a  remark about the role 
Regge symmetry in providing
a natural regularization on the representation ring
of $SU(2)$, once the set of spins  $a,b,c,d$ 
(and thus the  semi-perimeter $s$) is kept fixed.
If we denote $a$ the smallest among the eight (integer and semi-integer) 
parameters $(a,b,c,d,s-a,s-b,s-,s-d)$, then we assign the label $c$ 
to the opposite entry, \emph{cf.} the first $6j$ in \eqref{Rsym1}, and
$d$ is chosen as the biggest among the two remaining labels. 
The ordering and the range of $c$ which turn out to be
compatible with all of the quadrangular and triangular inequalities are given by 
\begin{equation}\label{Rsym2}
a \leq b \leq d \leq s\, ;\; d-(b-a) \leq c \leq d+ (b-a)\, 
\end{equation}
and $x_{max} -x_{min}=$ $y_{max} -y_{min}=2a$. 
It is easily checked that  the semi--perimeter  $s$ satisfies 
\begin{equation}\label{Rsym3}
s\, \leq \, \max \{ a,b,c,d\} \;\equiv\;  d+(b-a)\,, 
\end{equation}
a condition to be compared  with what would happen for the $SU(2)_q$
case with $q$ a  root of unity. The following inequality,
proven explicitly in \cite{Igor} and adapted here to our conventions on spin labeling,
\begin{equation}\label{Rsym4}
s\, \leq \, \min \{ a,b,c,d\} \,+\,\tfrac{(r-2)}{2}\,:=\, a +\kappa\,,
\end{equation}
involves the integer $r \geq 3$ with $q=e^{2\pi i/r}\,$.
The two inequalities above lead to the compatibility condition
\begin{equation}\label{Rsym5}
\kappa \,\leq\,x_{min} +y_{min}\;\;\;\; \text{or}\; r \,\leq\,(2x_{min}+1) +(2y_{min}+1)\,,
\end{equation}  
interpreted as associated with a sort of 
effective $q$-regularization. The integer $r$
is thus bounded by the sum of the dimensions of the $SU(2)$ {\em irreps} labeled $x_{min}$, $y_{min}$
and the geometrical content of Eq.\ref{Rsym5} relies on the 
identification of $x_{min}$, $y_{min}$ with the 
lower admissible values of the lengths of the diagonals 
of the quadrangle in Fig.\ref{App4} or of two opposite edges in
the solid Euclidean tetrahedron in
Fig.\ref{App5} of Appendix \ref{AppPrima}. It is worth noting that, owing to the nested structure of the quintuple of quadrangles, it can be easily shown that
the regularization conditions above can be fixed on a single reference
quadrangle (or even on a single tetrahedron in  $\mathcal{S}_{spin}$).

A quite interesting reformulation of the  
remarks presented so far
would be in term of a quaternionic reparametrization of the entries of the $6j$ 
as given in Eq.\ref{Rsym1}: the semi-perimeter $s$ represent the real part of a quaternion 
$\mathfrak{Q}$ and three independent linear combinations of $(a,b,c,d)$ give its
other real components. Then it turns out that that the Regge--transformed 
$6j$ is associated with the `conjugate' $\mathfrak{Q}'$ of quaternion $\mathfrak{Q}$,
a circumstance that justifies algebraically the use of term `Regge--conjugate'
for the $6j$ on the right side of Eq.\ref{Rsym1}
(note that octonions should also be called into play as  consistent coordinatization of pairs ($\mathfrak{Q}$, $\mathfrak{Q}'$)).
More details on such a reparametrization,  its implications and a few applications
can be found in \cite{AqMaMaJPA2013,Tesi}.

%%%%%%%%%%%%%%%%%%%%%%%%%%%%%%%%%%
%%%%%%%%%%%%%%%%%%%%%%%%%%%%%%%%%%
\section{\label{SecQuarta} Outlook and conclusions}
%%%%%%%%%%%%%%%%%%%%%%%%%%%%%%%%%%%%%%%%
%%%%%%%%%%%%%%%%%%%%%%%%%%%%%%%%%%%%%%%%%
Actually quite a lot of issues and interconnections have not been addressed here for lake of space. 
As for the BE identity, recall that it has been used originally for proving the combinatorial invariance of the Ponzano--Regge state sum under refinement, {\em cf.}  the moves depicted in Fig.\ref{App8} and Fig.\ref{App9}. In the same paper  it can be found  the derivation of the three-term recursion relation for the $6j$ symbol that in the semiclassical limit (where all angular momentum variables are much greater than 1 in $\hbar$ units) 
gives the asymptotic formula for the $6j$. As mentioned already in the introduction, this model has been extensively used in discrete quantum gravity approaches, as well as in topological quantum computing
see \cite{MaRa2002,MaRa2005,GaMaRa2009,KaMaRa2009}.
Of particular interest in quantum gravity applications is  the volume operator of the tetrahedron:
on the basis of the paper \cite{CaCaMa2002} we have developed  recently a complete analysis of the three-term recursion relation for the eigenfuntions of such operator highlightening the proper Regge--invariant frame of Racah quadratic algebra
\cite{Skly1982,GrLuZhAOP1992,AqMaMaJPCS2014}, as well as the emergence of a Hamiltonian dynamics at the classical level \cite{AqMaMaJPA2013}. 

In summary, we have emphasized in the present paper
the central role played in Regge--regularized Desargues spin networks  
by the BE identity (together with the orthogonality relation) as 
the common algebraic counterpart  of assemblings of
quintuples of projectively--related  quadrangles and tetrahedra, 
as well as of the associated geometric realizations. 
As for dimensional reduction 
through projections from 3D to 2D and 1D of Desargues spin networks, we argue that
it should be possible to proceed  in parallel with  semiclassical limiting procedures  achieved by taking suitable partial limits of the three--term recursions for the $6j$ symbol and for the volume
operator quoted above. The level of persistence of Regge symmetry might represent a guiding principle in this kind of analysis. 
Finally, as for the 4-simplex -- realized through  the space-dual of a Deasargues spin network-- 
we have shown in section \ref{SecSecondaC}   how its 3D cross sections still represents the (original)  Desargues spin network, thus opening the possibility of looking at dimensional reduction from 4D to 3D in a new perspective.
We argue also that, by properly framing the projective tools  within 
a picture grounded into  recursion relationships governed by discrete variables,   
a Hamiltonian description for the amplitude of the 4-simplex might emerge. Work is in progress in this direction.

%\vfill
%\newpage

\section{Appendix: Projective configurations and angular momentum theory}
 %%%%%%%%%%%%%%%%%%%%%%%%%%%%%%%%%%%%%%%%%%%%%%%%%%%%%%%%%%%%%%%%%%%%%%%%%%%
 %%%%%%%%%%%%%%%%%%%%%%%%%%%%%%%%%%%%%%%%%%%%%%%%%%%%%%%%%%%%%%%%%%%%%%%%%
 
 %%%%%%%%%%%%%%%%%%%%%%%%%%%%%%%%%%%%%%%%%%%%%%%%%%%%%%%%%%%%%%%%%%%%%%%%%%%%%%%%%%%%%
 \subsection{\label{AppPrima}The triangle of couplings and the quadrangle of recouplings}
\begin{figure}[h]
\begin{minipage}{11pc}
\includegraphics[width=11pc]{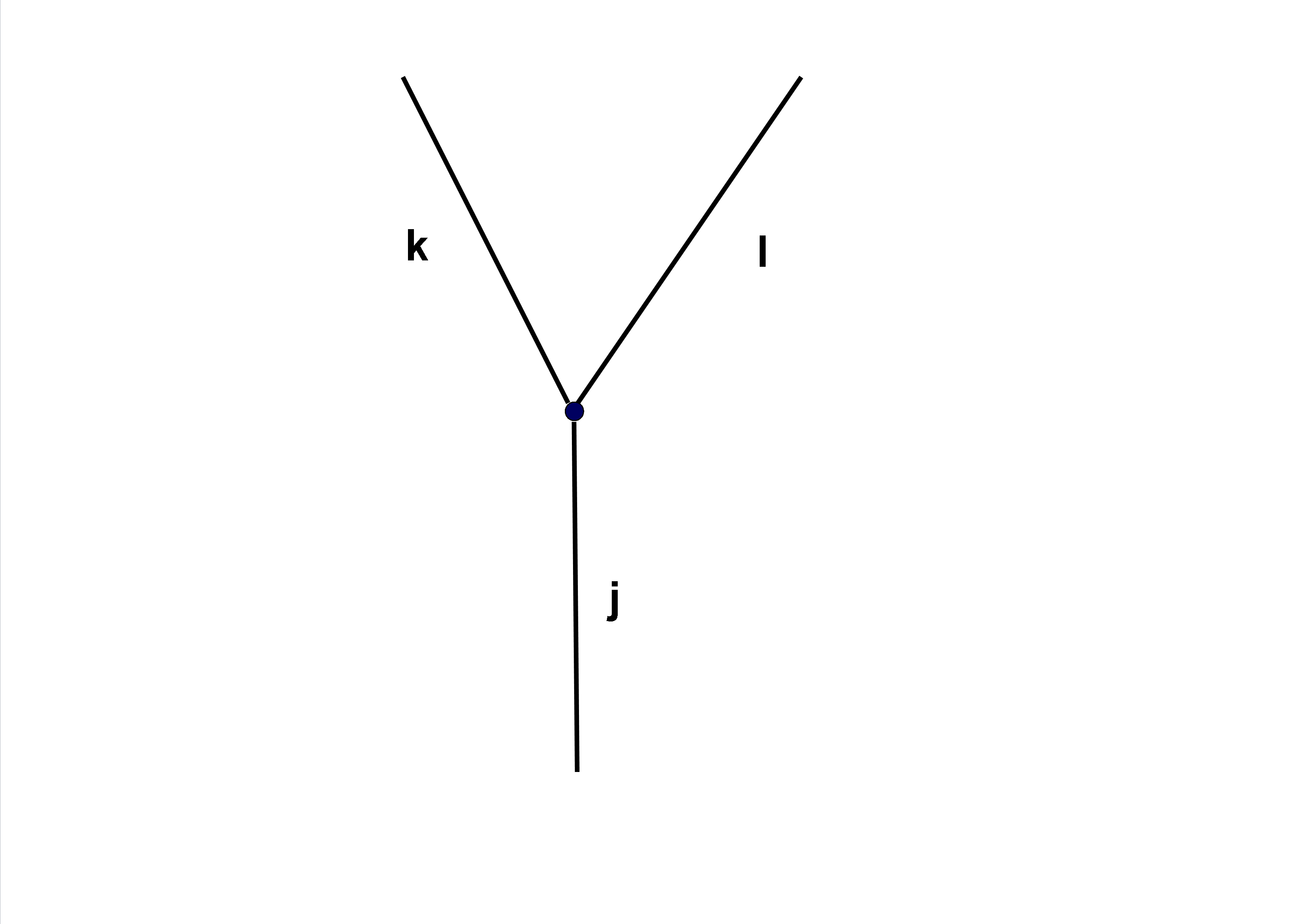} 
\caption{\label{App1} The triad $(j,k,l)$ associated with a trivalent vertex.}
\end{minipage}\hspace{2pc}%
\begin{minipage}{11pc}
\includegraphics[width=11pc]{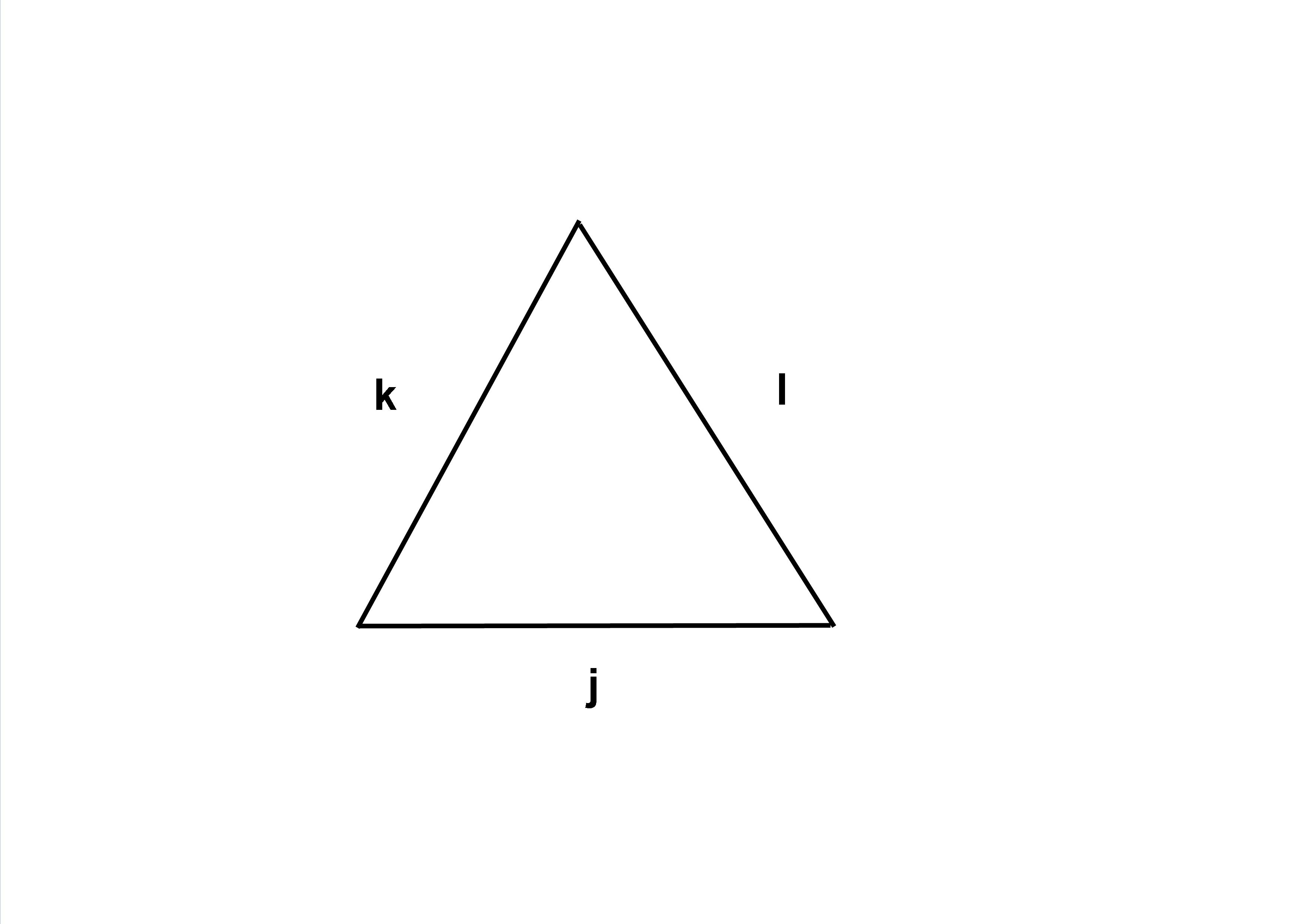}
\caption{\label{App2} The triad $(j,k,l)$ associated with a triangle.}
\end{minipage} \hspace{2pc}%
\begin{minipage}{11pc}
\includegraphics[width=11pc]{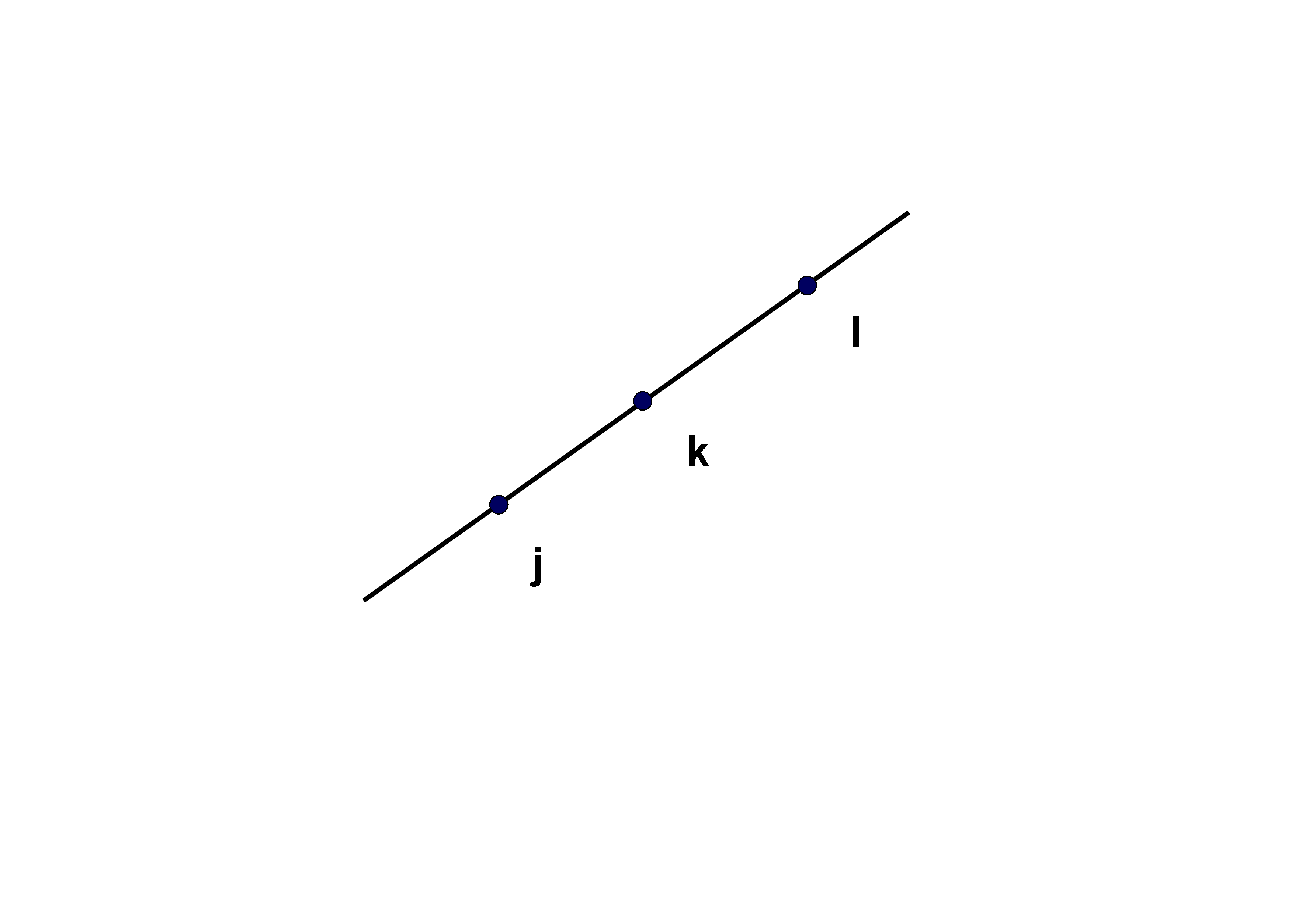}
\caption{\label{App3} The triad $(j,k,l)$ associated with three points on a same line.}
\end{minipage} 
\end{figure}

 %%%%%%%%%%%%%%%%%%%%%%%%%%%%%%%%%%%%%%%%%%%%%%%%%%%%%%%%%%%%%%%%%%%%%%%%%%%%%
 Triangular inequalities involving three {\em irreps} labels of $SU(2)$ angular momenta 
basically arise from
the Clebsch--Gordan series, $\mathcal{H}^{j_1} \otimes \mathcal{H}^{j_1}$ $ = \oplus_{j=|j_1-j_2|}^{j_1+j_2}$
$\mathcal{H}^j$, where $\mathcal{H}^j$ is the $(2j+1)$--dimensional Hilbert space of the (square of the) total angular momentum $\mathbf{J}$ $(-j \leq m \leq j)$ with $m$ being the magnetic quantum number associated with $J_z$,
the component of $\mathbf{J}$ along the quantization axis. The two independent systems of sharp angular momenta
$\mathbf{J}_1$ and $\mathbf{J}_2$ have dimensions satisfying $(2j_1+1)$ $(2j_2+1)$ $=(2j+1)$ and the (orthogonal) transformation between the basis vectors of $\mathcal{H}^{j_1}$ $\otimes \mathcal{H}^{j_1}$ and those of $\mathcal{H}^j$ is given by   Clebsch--Gordan coefficients or by their symmetrized counterpart, 
the Wigner $3j$ symbol
$\left( \begin{smallmatrix}
j_1 & j_2 &j\\
m_1 &m_2 & m
\end{smallmatrix}\right)$ with $m_1+m_2=m$. 
Such an unordered triple of angular momentum variables (spins), say $(j,k,l)$
taking values in $\{0,1/2,1,3/2,\dots\}$, constitutes a {\em triad} and can be graphically
represented  as either a trivalent vertex (Fig.\ref{App1}), or as a triangle (Fig.\ref{App2}),
or else by three points lying on a same line (Fig. \ref{App3}).
The latter was first noticed  by Fano and Racah \cite{FaRa1959},
hence unveiling the intriguing  connection between (binary) coupling of angular momenta and 
(finite) projective configurations of points and lines discussed below.

The basic building block of angular momentum theory is the $W$ recoupling coefficient of Racah or, equivalently,
 the Wigner $6j$ symbol ($a,b,c,d,x,y \in \{0,1/2,1,\dots \}$)
 \begin{equation}\label{6jApp}
 \begin{Bmatrix}
  a & b & x\\
c & d & y
 \end{Bmatrix}\;\;\text{with triads}\; (abx),\,(bcy),\,(cdx),\,(ady)
 \end{equation}
As is well known there exist three alternative graphical presentations of the 
combinatorial content of this
symbol, discussed below focusing on their interpretations in projective geometry.

\begin{figure}[h]
\begin{minipage}{11pc}
\includegraphics[width=11pc]{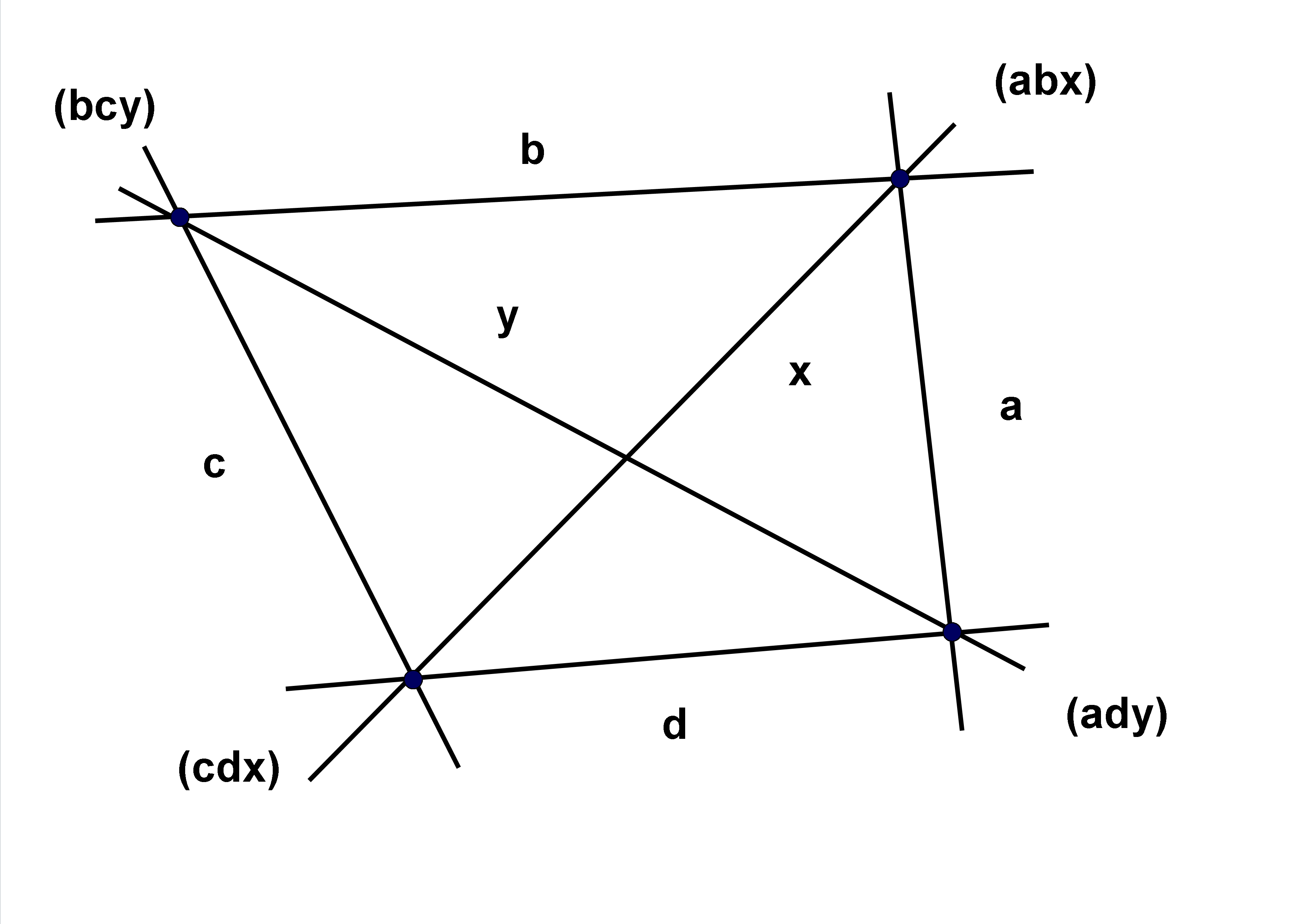}
\caption{\label{App4} The configuration $(4_3\,,6_2)$: four points in a plane are joined in pairs by six distinctlines.}
\end{minipage}\hspace{2pc}%
\begin{minipage}{11pc}
\includegraphics[width=11pc]{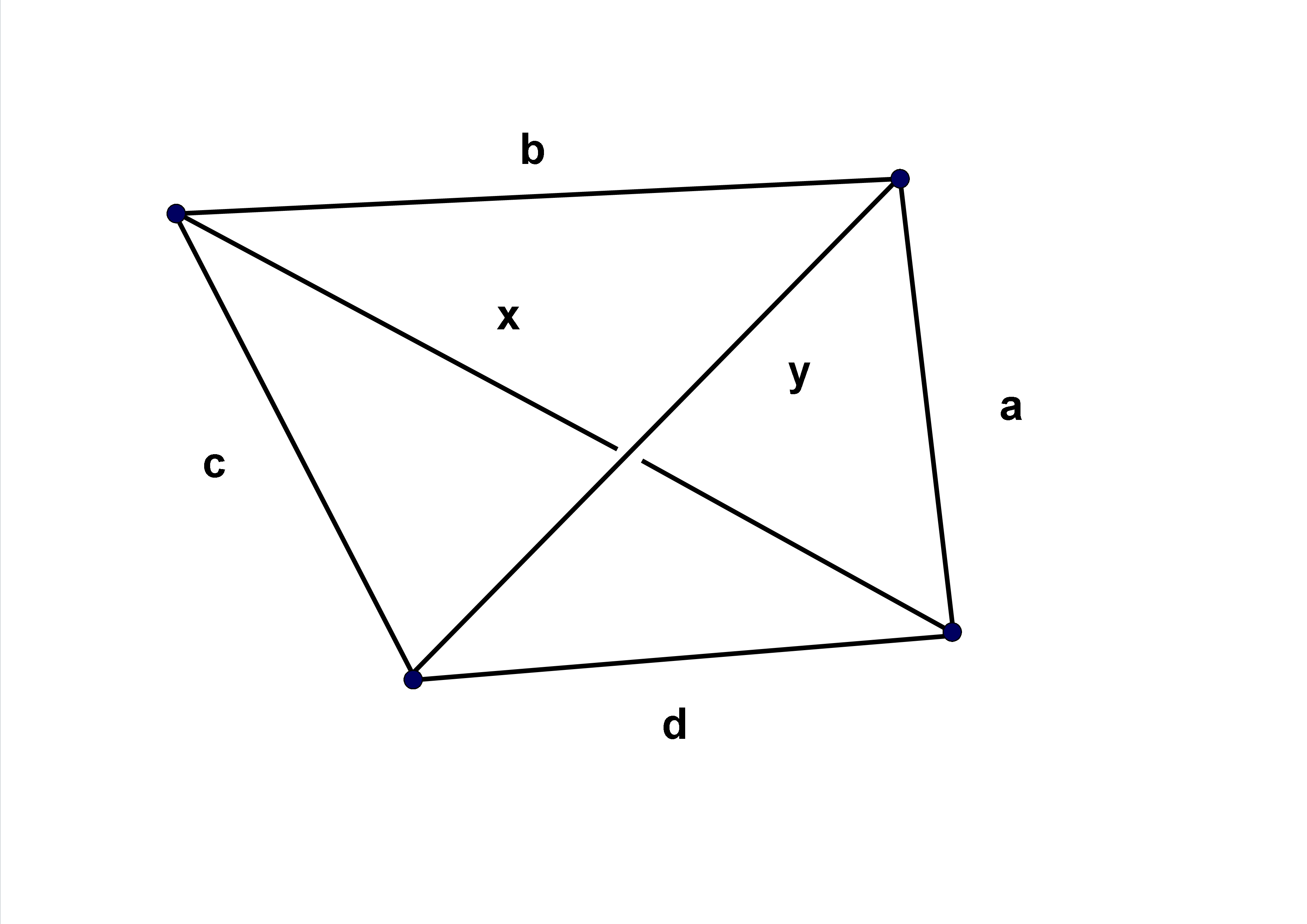}
\caption{\label{App5} The tetrahedron (the  boundary of a 3-simplex) has four vertices, 
six edges and four triangular faces.}
\end{minipage} \hspace{2pc}%
\begin{minipage}{11pc}
\includegraphics[width=11pc]{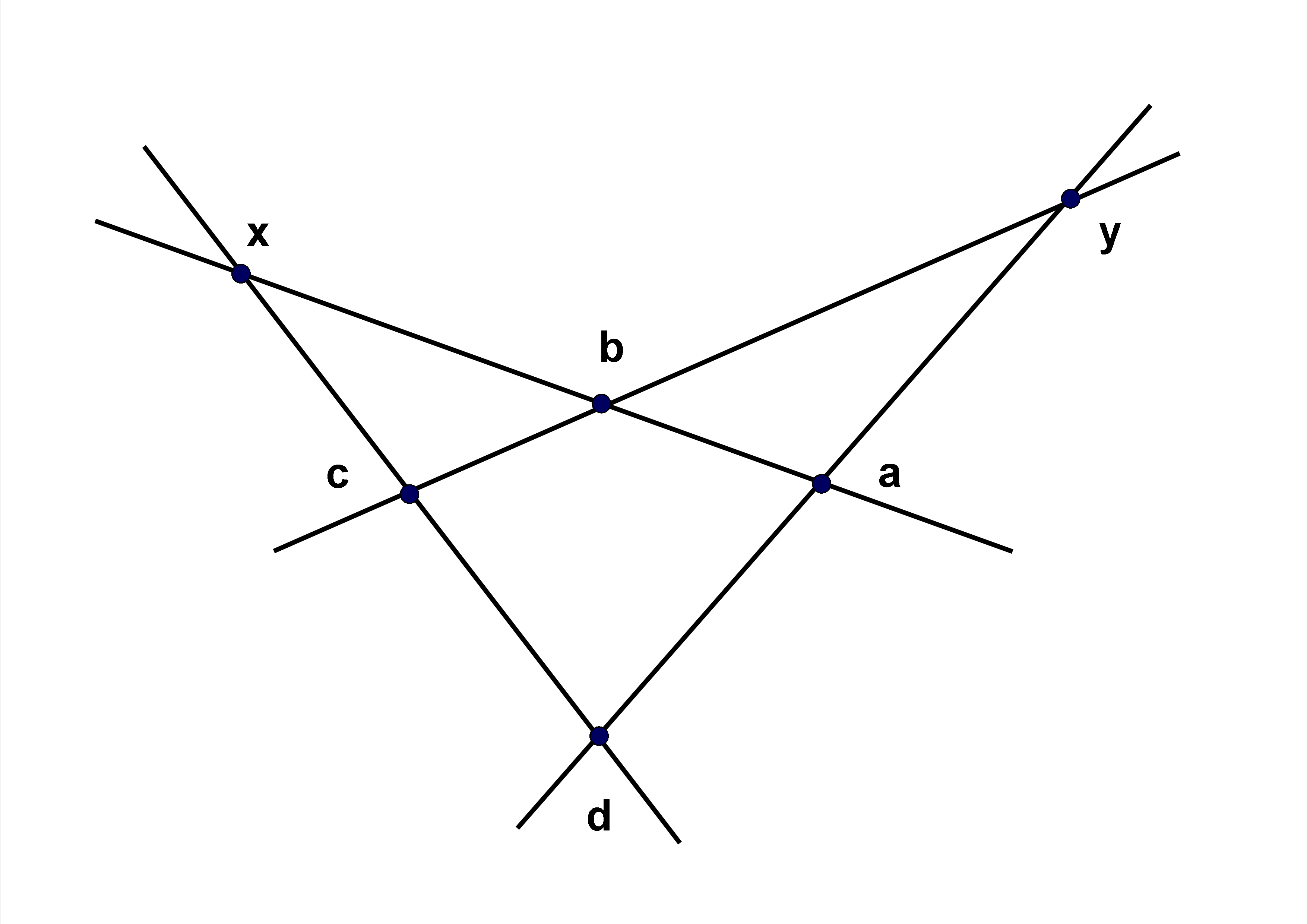}
\caption{\label{App6} The configuration $(6_2\,,4_3)$: four lines in a plane meet by pairs in six distinct points.}
\end{minipage} 
\end{figure}

{\bf The $6j$ as a complete quadrangle} (Fig.\ref{App4})\\
Edmonds \cite{Edm} showed that the $6j$ --together with other recoupling coefficients such
as the $9j$ and the $12j$ symbols-- can be associated with trivalent graphs obtained by joining the free ends of binary coupling trees  whose labeling are consistent with that of the trivalent vertex of Fig.\ref{App1}. 
On this basis Yutsis and  collaborators \cite{YuLeVa} elaborated a general diagrammatical scheme able to 
work out complicated  calculations --involving generalized Clebsch--Gordan coefficients and $3nj$
recoupling coefficients-- by resorting to combinatorial operations on such labeled graphs. An upgraded
version of these methods is collected in the handbook by Varshalovich and collaborators \cite{Russi},
the notation of which is used throughly in this paper. In particular,  the $6j$ is depicted as 
a complete graph on four vertices as in Fig.\ref{6jYutsis}, each vertex being
associated with a triad of spin labels. Notably, when each edge is extended to a line, the resulting 
configuration can be perceived projectively as a {\em complete quadrangle}, denoted $(4_3\,,6_2)$.\\
\noindent {\bf The $6j$ as a tetrahedron} (Fig.\ref{App5})\\
The second graphical presentation dates back to Wigner \cite{Wigner1959} and Ponzano and Regge \cite{PoRe}
 and has been especially exploited in connection with the semiclassical limit of the $6j$ symbol
 representing an Euclidean tetrahedron in the classically allowed region where its square
volume is positive. Here spin labels are associated with the edge lengths and the triads with the 
triangular faces as in Fig.\ref{App2}. 
The relationship between the two presentations  --quadrangle and tetrahedron-- is easily recognized 
in polyhedral geometry: if the quadrangle of Fig.\ref{App4} is realized in  Euclidean 3-space
as a  tetrahedron (with the condition on the volume given above), or, equivalently, 
if it represents the projection on a plane of a solid tetrahedron, then under topological duality of 2D 
simplicial complexes, (vertex $\leftrightarrow$ triangle) and (edge $\leftrightarrow$ edge), the two diagrammatical presentations are turned into each other.
(Obviously this correspondence involves the triangulation of the boundary of the tetrahedron 
while in the study of 3D simplicial dissections of manifolds  most of the results and applications
 rely on the duality (0-simplex $\leftrightarrow$ 3-simplex) and 
(1-simplex $\leftrightarrow$ 2-simplex)).
 
{\bf The $6j$ as a complete quadrilateral} (Fig.\ref{App6})\\
This presentation, derived from the assignment (triad $\leftrightarrow$ three points on a line) 
of Fig.\ref{App3}, has been  the basic tool to disclose the projective
content of angular momentum recoupling theory since Fano and Racah. We
refer the reader interested in more details to the papers of Robinson 
\cite{deRob1970}, Judd \cite{Judd1983},  Labarthe \cite{Lab2000} and to 
the book of Biedenharn and Louck \cite{BiLouck9}.  The quadrangle 
of Fig.\ref{App4} is related to the quadrilateral by the standard projective duality
in the plane, (point $\leftrightarrow$ line).

 It is worth to remark that  it is actually the
 configuration $(7_3)$ of seven points and seven lines
(known as the Fano plane) that historically has attracted much attention, and
has been also the object of recent investigations by the authors
in connection with applications to atomic and molecular physics, {\em cf.} 
\cite{Aquila6}
and references therein.

%%%%%%%%%%%%%%%%%%%%%%%%%%%%%%%%%%%%%%%%%%%%%%%%%%%%%%%%%
\subsection{\label{AppSeconda}Assembling of projective configurations: from quadrangles to Desargues configuration}
%%%%%%%%%%%%%%%%%%%%%%%%%%%%%%%%%%%%%%%%%%%%%%%%%%%%%%%%%%%%%%%%%%%%%%%%%

The three presentations of the $6j$ symbol  \eqref{6jApp}  --two of which are truly projective configurations as recognized above-- have been used 
alternately in different approaches to quantum spin networks, even though from the combinatorial viewpoint  they are related by simple (projective or Poincar\'e) duality transformations as pointed out above. Taken for granted that each of them matches  with the standard algebraic setting (in particular with the three-term recursion relation for the $6j$ \cite{Russi}), 
we are implementing 
a bottom--up  approach to construct spin networks from 
the basic axioms of projective geometry (as stated in \cite{Cox}) and showing how
 assembling  quadrangles  --rather than  quadilaterals-- turns out to be the
most economical and consistent way of looking at emergent geometric realizations of interest  also in discretized quantum gravity models, {\em cf.} the final remarks at the end of section \ref{SecSecondaC}.
The first two incidence axioms  are\\
{\bf 
I) Any two distinct points are incident with just a line.\\
II) Any two lines are incident with at least one point.}\\
A configuration in the plane is denoted  $(p_{\gamma}, \ell_{\pi})$, where $p$ is the number of
points, $\ell$ the number of lines, $\gamma$ the number of lines  per point, and $\pi$
the number of points per line. The equality $p \gamma=\ell \pi$ is easily verified, and 
configurations for which  $p =\ell$ and $\gamma= \pi$, so that $(p_{\gamma}, p_{\gamma})$
$\equiv$ $(p_{\gamma})$,  are called symmetric (for instance $(3_2)$ is a triangle and 
$(n_2)$  a polygon with $n$ sides). The (standard) projective dual  to a configuration
 $(p_{\gamma}, \ell_{\pi})$ is a configuration $(\ell_{\pi}, p_{\gamma})$ in which the roles
 of points and lines are interchanged, and symmetric configurations are self--dual 
in the plane by definition.
The complete quadrangle $(4_3, 6_2)$ of Fig.\ref{App4} and  quadilateral 
$(6_2, 4_3)$ of Fig.\ref{App6} are dual to each other, as already noted. 
The existence of the quadrangle bears on the axiom\\ 
{\bf III) There exist four points, no three of which are collinear.}\\
Actually this is the simplest configuration that allows for the assignment of 
coordinates in the plane, looking at the complete quadrangle $(4_3\,,6_2)$ of Fig.
\ref{App4}
with vertices  labeled by the pairs
$(0,0)$ $(0,1)$ $(1,0)$ $(1,1)$ taking values in the smallest finite number field $\mathbb{F}_2$
\cite{Hall1943}. 

\begin{figure}[hb]
\centering
\includegraphics[scale=0.20]{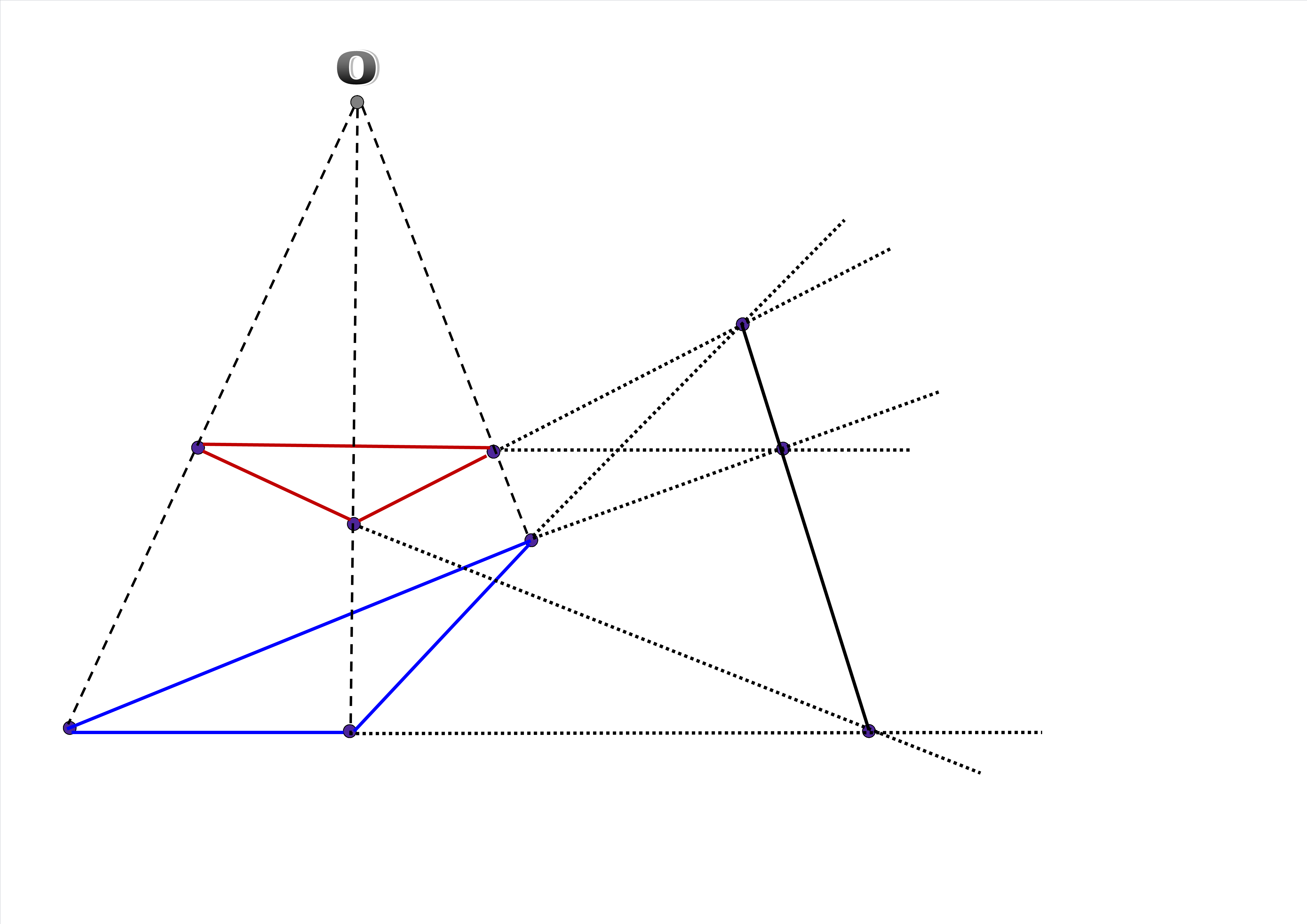}
\caption{\label{App7} Illustration of Desargues' theorem: if two triangles (red and blue) are in perspective from a center O, then the extensions to lines of their sides meet in a line, the perspective line. In the (plane) dual of this theorem  the existence of the perspective line with three points is assumed and the existence of the center of perspective O is proven.}
\end{figure}

Focusing on symmetric configurations, it should be clear that the 
the symbolic  expression $(p_{\gamma})$ does not determine uniquely a projective configuration up to incidence isomorphisms. In particular, among
the ten different $(10_3)$ configurations, the Desargues configuration plays 
a prominent role in projective geometry essentially because it encodes Desargue's theorem. The
statement is illustrated in Fig.\ref{App7}
Actually Desargues' theorem holds true in any projective geometry $P^n$ for $n \geq 3$ while  its 
realization in the plane $P^2$  must be separately postulated ({\em cf.} \cite{Cox}). 
The principle that characterizes projective geometry and is crucial for the hierarchical construction of spin networks is {\em duality} (the notion of plane duality, or duality in $P^2$, has been already quoted above). In general, suppose that in a proposition  --or a figure, or a configuration-- in $P^n$ one interchanges $P^n$ and
$P^{n-r-1}$ ($0 \leq r \leq n)$ and switches  `is contained'/ `contains',  `collinear'/ `concurrent'
and similar. Then the duality principle can be stated as follows\\\
 {\bf Duality principle.} If a proposition (about a configuration) is true in $P^n$ then it is also true in {\em any of the dual configurations}  of the original one lying in $P^{n-r-1}$ for $0 \leq r \leq n$.

\end{document}